\documentclass[final,1p,times]{elsarticle}
\usepackage{graphicx}
\usepackage{amsmath}
\usepackage[version=4]{mhchem}
\usepackage{float}
\usepackage{color}
\usepackage{caption}
\usepackage{subcaption}
\usepackage{tabularx}
\usepackage{comment}
\usepackage{breqn}
\usepackage{mathtools}
\usepackage[noabbrev]{cleveref}

\usepackage[linesnumbered,ruled,lined]{algorithm2e}

\journal{Journal of Computational Physics}

\begin{document}

\begin{frontmatter}






\title{Stochastic Operator Learning for Chemistry in Non-Equilibrium Flows} 


\author[label1]{Mridula Kuppa\fnref{label2}}
\fntext[label2]{Graduate Research Assistant, Department of Aerospace Engineering}
\author[label3]{Roger Ghanem\fnref{label4}}
\fntext[label4]{Professor, Department of Civil and Environmental Engineering, Department of Aerospace and Mechanical Engineering}
\author[label5]{Marco Panesi\corref{cor1}\fnref{label6}}
\cortext[cor1]{Corresponding author. E-mail address: mpanesi@illinois.edu}
\fntext[label6]{Professor, Department of Aerospace Engineering}

\affiliation[label1]{organization={University of Illinois at Urbana-Champaign},
city={Urbana}, 
state={IL},
postcode={61801},
country={USA}}

\affiliation[label3]{organization={University of Southern California},
city={Los Angeles}, 
state={California},
postcode={90089},
country={USA}}

\affiliation[label5]{organization={University of Illinois at Urbana-Champaign},
city={Urbana}, 
state={IL},
postcode={61801},
country={USA}}
            
\begin{abstract}
This work presents a novel framework for physically consistent model error characterization and operator learning for reduced-order models of non-equilibrium chemical kinetics. By leveraging the Bayesian framework, we identify and infer sources of model error and parametric uncertainty within the Coarse-Graining Methodology (CGM) across a range of initial conditions. The model error is embedded into the chemical kinetics model to ensure that its propagation to quantities of interest remains physically consistent. 
For operator learning, we develop a methodology that separates time dynamics from the parameters governing initial conditions, model error, and parametric uncertainty. Karhunen-Lo\`{e}ve Expansion (KLE) is employed to capture time dynamics, yielding temporal modes, while Polynomial Chaos Expansion (PCE) is subsequently used to map model error and input parameters to the KLE coefficients.
This proposed model offers three significant advantages: i) Separating time dynamics from other inputs ensures the stability of the chemistry surrogate when coupled with fluid solvers; ii) The framework fully accounts for model and parametric uncertainty, enabling robust probabilistic predictions; iii) The surrogate model is highly interpretable, with visualizable time modes and a PCE component that facilitates the analytical calculation of sensitivity indices, allowing for the ranking of input parameter influence.
We apply this framework to the $\mathrm{O}_2$-$\mathrm{O}$ chemistry system under hypersonic flight conditions, validating it in both a 0-D adiabatic reactor and coupled simulations with a fluid solver in a 1-D normal shock test case. Results demonstrate that the surrogate is stable during time integration, delivers physically consistent probabilistic predictions accounting for both model and parametric uncertainty, and achieves a maximum relative error below 10\%. This work represents a significant step forward in enabling probabilistic predictions of non-equilibrium chemical kinetics within coupled fluid solvers, offering a physically accurate approach for hypersonic flow predictions.  

\end{abstract}




\end{frontmatter}


\section{Introduction}\label{sec:Introduction}

Studying the aerothermal environment around hypersonic vehicles is a complex multiscale and multiphysics challenge, involving the interplay of fluid dynamics, chemical kinetics, and radiation across a broad range of length and time scales \cite{gnoffo1999planetary, park1989nonequilibrium}. As a spacecraft enters a planetary atmosphere, a strong bow shock forms in front of the vehicle, creating a high-temperature region. In this region, the gas species' internal energy modes become excited, leading to energy redistribution through collisions and resulting in changes in chemical composition due to dissociation and ionization reactions \cite{surzhikov2012radiative}. These excitation, transfer, and reaction timescales are comparable to the flow timescale, causing thermochemical nonequilibrium \cite{munafo2014boltzmann, aliat2003nonequilibrium}. Consequently, the internal energy distribution deviates significantly from the equilibrium Maxwell-Boltzmann distribution \cite{panesi2014nonequilibrium}. Accurate characterization of this non-Boltzmann distribution is crucial, as it directly affects convective heat transfer, radiation, and the design of the spacecraft's thermal protection system.

The most accurate description of the non-Boltzmann distribution is achieved through the rovibrational collisional model \cite{panesi2013rovibrational, schwenke1990theoretical, gang2012state}. In this model, each internal energy level is treated as a separate species, and the population of each level is obtained by solving master equations. However, with thousands of possible rovibrational states (about 6500 for $\mathrm{O}_2$ and about 9300 for $\mathrm{N}_2$), this approach involves millions of reactions with multiquantum jumps, making it computationally expensive. Consequently, it is typically limited to 0-D and 1-D simulations and cannot be applied to multidimensional CFD simulations \cite{panesi2013rovibrational, panesi2014nonequilibrium, munafo2013modeling, macdonald2016nonequilibrium}. Additionally, the stiffness of these equations further increases computational costs.

 Reduced-order models based on Coarse Graining Methodology (CGM) provide an accurate characterization of the non-Boltzmann distribution at a significantly lower computational cost compared to state-to-state (STS) models in multidimensional CFD simulations \cite{liu2010multi, magin2012coarse, liu2015general, munafo2020reduced, sharma2019application}.
 In CGM, internal energy levels are grouped into bins, treating each bin as a pseudo-species. This approach reduces the number of chemical reactions, making the calculation of the chemistry source term more manageable. The CGM involves choosing a grouping strategy and a reconstruction strategy, which serves as a closure model for the macroscopic governing equations derived by summing the moments of the master equations based on the chosen grouping strategy.

 Various grouping strategies exist, including spectral clustering, distance from the centrifugal barrier, and energy-based binning \cite{liu2015general, sahai2017adaptive, venturi2020data,jacobsen2024information}. These methods use deterministic grouping strategies by hard-assigning energy levels to groups. The most common reconstruction strategy for deriving STS populations from group populations is based on the maximum entropy principle \cite{heims1963moment, levermore1996moment}. This principle represents energy-state populations as an exponential function of linear combinations of monomial basis functions of state energies. In practice, this expansion is truncated at the constant, linear, or quadratic term \cite{sharma2019characterization, sharma2020coarse}. The coefficients of these basis functions are determined from moment constraints. Increasing the number of basis terms raises computational costs due to additional constraint equations. When truncated to the linear term, reconstruction can use either the translational temperature (solving only zeroth moment equations) or ``group internal temperatures" (extracted from group internal energies obtained by solving the first moment of master equations).

Several sources of uncertainty and error in CGM can be identified:
\begin{enumerate}
\item Uncertainty in grouping strategy: Spectral clustering for excitation dynamics and centrifugal barrier-based grouping for dissociation have produced accurate results, but new chemistry systems might lack state-specific rate coefficient information, making grouping strategy uncertain. Energy-based binning, while simpler, fails to capture mode-specific dynamics, introducing additional model error \cite{venturi2019machine}.
\item Truncation in reconstruction strategy: Maximum entropy principle reconstructions are truncated, leading to model error due to neglected higher-order terms. This error is challenging to quantify. Using translational temperature rather than group temperature in linear reconstruction increases error as it does not satisfy internal energy constraints. Adding more groups can reduce this error but increases computational costs.
\item Uncertainty in chemical kinetic rate coefficients: Prediction uncertainty in high-temperature flow simulations arises from uncertainties in Potential Energy Surface (PES) fitting, cross-section determination from Quasi Classical Trajectory calculations (QCT), and rate coefficient fitting 
\citep{west2014multistep,west2015uncertainty,lockwood2010uncertainty,venturi2020bayesian,wang2015combustion,zhao2019sensitivities}.
\end{enumerate}
These factors highlight the need for a framework to quantify model error and parametric uncertainty in reduced-order models for non-equilibrium chemistry to achieve reliable predictions with meaningful uncertainties.

Model error quantification, also known as structural or model inadequacy, is a growing area of research with two main approaches - explicit statistical and implicit statsictical corrections. The first involves adding statistical terms to specific model outputs to account for the bias between model predictions and observations. 
The foundational work by Kennedy and O'Hagan \cite{kennedy2001bayesian} introduced this approach, but it has been noted that it may potentially violate physical laws, struggle with error extrapolation to other quantities of interest (QoIs), and sometimes conflate bias correction with measurement noise \cite{oliver2015validating,bayarri2007framework,pernot2017parameter}. Extensions of this method have been developed to include all relevant inputs, improving its ability to extrapolate to scenarios outside calibration \cite{xu2015bayesian,neufcourt2018bayesian}. The second approach, more recent and adopted in this work, embeds stochastic terms within model components, propagating them downstream to all QoIs. This methodology respects physical constraints, allows for extrapolation to unobserved QoIs, and identifies areas of the model needing further refinement. It has been successfully applied in various fields, including chemical ignition modeling \cite{sargsyan2015statistical}, health economics \cite{strong2012managing}, RANS and LES simulations \cite{oliver2011bayesian,emory2011modeling,huan2017global}, and high-speed Lagrangian flow simulations \cite{kuppa2024model}.  

Machine learning is increasingly being used to accelerate the calculation of chemistry source terms, thereby speeding up reactive flow CFD simulations \cite{bielawski2023highly,barwey2021neural,leal2020accelerating,mao2023deepflame}. Campoli et al. \cite{campoli2022assessment} compared several ML strategies for regressing chemistry source terms and coupling these models with ODE solvers, finding that performance gains depend on the interfaced codes, with greater benefits when the source term calculation is more computationally expensive. Ozbenli et al. \cite{ozbenli2020numerical} demonstrated a threefold speed-up using an artificial neural network as a surrogate for a vibrational-specific state-to-state model compared to a conventional time integrator. Zanardi et al. \cite{zanardi2023adaptive} developed a deep-learning framework for reduced-order rovibrational models, employing it in CFD simulations using an operator splitting approach. While these studies focus on deterministic surrogate modeling for computational speedup, they do not address parametric uncertainty or model error and often lack interpretability due to the black-box nature of machine learning frameworks.  

In this work, we employ a Bayesian framework to characterize model error in CGM due to uncertainty in the grouping strategy. We focus on a 2-bin model that divides energy levels into two groups, embedding model error in the energy partition by expanding it as a first-order Polynomial Chaos Expansion (PCE). The PCE coefficients are inferred using Bayesian inference, and we also account for uncertainty in grouped rate coefficients stemming from uncertainty in state-specific rate coefficients. The posterior distributions of energy partition and grouped rate coefficients are subsequently propagated to the QoIs through a surrogate model that approximates the solution operator of this stochastic 2-bin model. The surrogate model decouples time dynamics from the influence of other input parameters. This is achieved by first using Karhunen-Loeve Expansion (KLE) to learn the time dynamics and then employing PCE to map initial conditions to the KLE coefficient space. The constructed surrogate model is applied in time-marching simulations for 0-D adiabatic reactors and 1-D normal shocks, where it is coupled with a flow solver via operator splitting. 
The surrogate is probabilistic, accounting for model error and parametric uncertainty, and is easier to integrate with legacy flow solvers compared to machine learning models, while retaining interpretability. 

The paper is organized as follows: Section \ref{sec:Physical modelling} covers the governing equations for flows with thermochemical non-equilibrium, the coarse grain methodology, and the equations for a 0-D adiabatic reactor, along with the governing equations for time-accurate simulation of a 1-D normal shock. In Section \ref{sec:Uncertainty quantification}, we discuss the framework developed for quantifying model and parametric uncertainty, as well as the surrogate modeling methodology. Results are presented in Section \ref{sec:Results and Discussions}, followed by conclusions in Section \ref{sec:Conclusions}.

\section{Physical Modelling}\label{sec:Physical modelling}

The governing equations for chemically reacting flows with state-to-state modeling of thermo-chemical non-equilibrium include the species conservation equation for each internal energy level, as well as the momentum and energy conservation equations. These can be expressed as:
\begin{align}
\frac{\partial \rho_i}{\partial t} + \nabla \cdot \left(\rho_i (\mathbf{v} + \mathbf{U}_i)\right) &= \omega_i \quad i \in I_b \label{eq:NS_spe}\\
\frac{\partial \rho \mathbf{v}}{\partial t} + \nabla \left(\rho \mathbf{v}\mathbf{v} + p I\right) &= \nabla \tau \label{eq:NS_mom}\\
\frac{\partial \rho E}{\partial t} + \nabla \cdot \left(\rho H \mathbf{v}\right) &= \nabla \cdot \left(\tau \mathbf{v} - \mathbf{q}\right) \label{eq:NS_eng}
\end{align}
Here, $t$ represents time, $\rho$ denotes the total mass density, $\rho_i$ is the partial density of rovibrational energy state $i$, and $I_b$ denotes the set of bound energy levels. $\mathbf{v}$ and $\mathbf{U}_i$ denote mass averaged velocity and diffusion velocity respectively. $p$ is the pressure and $I$ is the identity tensor. $\omega_i$ is the mass production source term due to collisional and radiative processes. $\tau$ denotes the stress tensor and $\mathbf{q}$ is the heat flux vector. $E$ and $H$ denote the total specific energy and enthalpy respectively. The computation of kinetic source terms is detailed below, while other transport and thermodynamic properties are discussed in \cite{munafo2023plato,munafo2020computational}. 

In this work, we study the $\mathrm{O}_2$-$\mathrm{O}$ chemistry system. Assuming all atoms and molecules to be in their electronic ground state, the set of chemical processes considered in this work can be written as:
\begin{align}
\text{Excitation and de-excitation:} \quad \mathrm{O}_2(i)+\mathrm{O} & \xrightleftharpoons[k_{ji}]{k_{ij}} \mathrm{O}_2(j) +\mathrm{O} \label{eq:sts_excit}\\
\text{Dissociation and recombination:} \quad 
\mathrm{O}_2(i)+\mathrm{O} & \xrightleftharpoons[k_{i}^r]{k_{i}^d} \mathrm{O}+\mathrm{O}+\mathrm{O} \label{eq:sts_diss}
\end{align}
Here, 
$i$ and $j$ represent distinct energy levels within $I_b$; $k_{ij}$ and $k_{ji}$ are the excitation and de-excitation rate coefficients, respectively between levels $i$ and $j$, while $k^d_i$ and $k^r_i$ are the dissociation and recombination rate coefficients for level $i$. These coefficients satisfy the micro-reversibility condition and are typically derived from QCT calculations on fitted PESs, both of which introduce uncertainties. Additionally, these rate coefficients are often fitted using a modified Arrhenius function of translational temperature, further contributing to uncertainty.

The source terms in species conservation equation $\omega_i$ can now be expressed as:
\begin{align}
    \omega_i = m_{\mathrm{O}_2}\sum_{j \in I_b} \left[-k_{ij}n_i n_{\mathrm{O}} + k_{ji} n_j n_{\mathrm{O}}\right] + \left[-k_i^d n_i n_{\mathrm{O}} + k_i^r n^2_{\mathrm{O}}n_{\mathrm{O}}\right]
\end{align}
Here, $m_{\mathrm{O}_2}$ is the molecular mass of $\mathrm{O}_2$ species, $n_i$ denotes the number density of level $i$, and $n_{\mathrm{O}}$ represents the number density of $\mathrm{O}$ species.

State-to-state modeling is computationally prohibitive due to the large number of internal energy states involved. This necessitates the use of a reduced-order model to study thermo-chemical non-equilibrium. In this work, we employ the Coarse-graining methodology, which will be detailed in the next section.
    
\subsection{Coarse-graining methodology}

In this approach, internal energy levels are grouped into bins, and the dynamics of bin-specific properties like population and internal energy are then solved. The method involves three key steps:
\begin{enumerate}
    \item Deciding the binning strategy
    \item Choosing a bin-wise distribution function to reconstruct state-specific population
    \item Deriving the macroscopic governing equations based on the chosen bin-wise distribution function
\end{enumerate}
Various binning strategies have been developed, including energy-based binning, where bins are non-overlapping continuous intervals of equal width in the internal energy space; adaptive binning, which considers state-specific rate coefficients to determine bins; and vibrational-specific binning, where levels with the same vibrational quantum number are grouped together \cite{liu2015general,sahai2017adaptive,venturi2020data}.
In this work, we use a linear bin-wise distribution function that maximizes entropy for state-specific reconstruction. For a group $g$, this is expressed as:
\begin{equation}
    \mathcal{F}_g (\epsilon_i): \quad \log \left(\frac{g_i}{n_i}\right) = \alpha_g + \beta_g \epsilon_i \label{eq:grp_recon}
\end{equation}
where $n_i$, $g_i$, and $\epsilon_i$ represent the population, degeneracy and internal energy of state $i$ in group $g$. The bin-specific constants $\alpha_g$ and $\beta_g$ are determined by the bin population and energy constraints:
\begin{align}
    n_g &= \sum_{i \in I_g} n_i \label{eq:grp_pop}\\
    e_g &= \sum_{i \in I_g} n_i \epsilon_i \label{eq:grp_eng}
\end{align}
where $I_g$ denotes the set of states in group $g$, $n_g$ is the total number density of group $g$, and $e_g$ is the total group internal energy.
For convenience, internal temperature for each group can be defined as:
\begin{equation}
    T_g = \frac{1}{k_B \beta_g}
\end{equation}
where $k_B$ is the Boltzmann constant. 
Taking the zero-th and first order moments of \cref{eq:NS_spe} with $\epsilon_i$ as the weight yields the bin-specific master equations:
\begin{align}
    \sum_{i \in I_g} \frac{\partial \rho_i}{\partial t} + \nabla \cdot \left(\rho_i (\mathbf{v} + \mathbf{U}_i)\right) &= \omega_g\label{eq:zero_mom}\\
    \sum_{i \in I_g} \left(\frac{\partial \rho_i}{\partial t} + \nabla \cdot \left(\rho_i (\mathbf{v} + \mathbf{U}_i)\right) \right) \epsilon_i &= \Omega_g \label{eq:first_mom}
\end{align}
The source terms on the right-hand side can be conveniently represented in terms of grouped rate coefficients as: 
\begin{align}
    \omega_g &= \sum_{i \in I_g} \omega_i = m_{\mathrm{O}_2}\left( \sum_{h \in \mathbb{P}} \left[-{}^0K_{gh}n_g n_{\mathrm{O}} + {}^0K_{hg}n_h n_{\mathrm{O}}\right] - {}^0C_g^dn_gn_{\mathrm{O}} + {}^0C_g^rn^2_{\mathrm{O}}n_{\mathrm{O}}\right)\\
    \Omega_g &= \sum_{i \in I_g} \omega_i \epsilon_i = m_{\mathrm{O}_2}\left( \sum_{h \in \mathbb{P}} \left[-{}^1K_{gh}n_g n_{\mathrm{O}} + {}^1K_{hg}n_h n_{\mathrm{O}}\right] - {}^1C_g^dn_gn_{\mathrm{O}} + {}^1C_g^rn^2_{\mathrm{O}}n_{\mathrm{O}}\right)
\end{align}
Here, $\mathbb{P}$ represents the set of all groups. The grouped rate coefficients are derived from state-specific rate coefficients as:
\begin{align}
    {}^mK_{gh} &= \sum_{i \in I_g} \sum_{j \in I_h} k_{ij}\epsilon_i^m n_i/n_g = \sum_{i \in I_g} \sum_{j \in I_h} k_{ij}\epsilon_i^m g_i \exp{\left(-\epsilon_i/\left(k_B T_g\right)\right)}/Q_g\label{eq:macro_rate_gh}\\
    {}^mK_{hg} &= \sum_{i \in I_g} \sum_{j \in I_h} k_{ji}\epsilon_i^mn_j/n_h = \sum_{i \in I_g} \sum_{j \in I_h} k_{ji}\epsilon_i^m g_j \exp{\left(-\epsilon_j/\left(k_B T_h\right)\right)}/Q_h \label{eq:macro_rate_hg}\\
    {}^mC_g^d &= \sum_{i \in I_g} k_i^d\epsilon_i^m n_i/n_g = \sum_{i \in I_g} k_i^d\epsilon_i^m g_i \exp{\left(-\epsilon_i/\left(k_B T_g\right)\right)}/Q_g\label{eq:macro_rate_g}\\
    {}^mC_g^r &= \sum_{i \in I_g} k_i^r\epsilon_i^m \label{eq:macro_rate_h}
\end{align}
with $m = 0$ or $1$. In \crefrange{eq:macro_rate_gh}{eq:macro_rate_g}, the second equality is derived using \crefrange{eq:grp_recon}{eq:grp_eng} along with the definition of the group partition function: $Q_g = \sum_{i \in I_g} g_i \exp\left({-\epsilon_i/\left(k_B T_g\right)}\right)$. For a detailed derivation, see \cite{liu2010multi,liu2015general,sharma2020coarse}. It is crucial to note that while grouped rate coefficients for mass (m=0) and energy (m=1) are different, they are not independent; thus, uncertainty in state-specific rate coefficients propagates to these grouped rate coefficients in a coupled manner. This approach reduces the number of governing equations from over 6000 (for the $\mathrm{O}_2$-$\mathrm{O}$ system) to the order of the number of groups. 

\subsection{Zero dimensional adiabatic reactor}

In a 0-D reactor, the coarse-grained \crefrange{eq:zero_mom}{eq:first_mom} along with the total energy \cref{eq:NS_eng} reduce to:
\begin{align}
    \frac{\partial \rho_g}{\partial t} &= \omega_g \label{eq:0D_adia_species}\\
    \frac{\partial \rho_g \Tilde{e}_g}{\partial t} &= \Omega_g \label{eq:0D_adia_int_eng}\\
    \frac{\partial \rho E}{\partial t} &= 0 \label{eq:0D_adia_tot_eng}
\end{align}
where $\Tilde{e}_g$ denotes the average group internal energy. 
These equations govern an adiabatic thermodynamic system with no external mass or energy exchange. 
For numerical solution in practice, \crefrange{eq:0D_adia_species}{eq:0D_adia_tot_eng} are rearranged in terms of species mass fractions and temperatures \cite{munafo2023plato}. 

\subsection{One dimensional normal shock}\label{sec:shock_goveq}
In this scenario, assuming inviscid flow, the coarse-grained moment \crefrange{eq:zero_mom}{eq:first_mom} along with the momentum \cref{eq:NS_mom} and energy \cref{eq:NS_eng} simplify to Euler equations:
\begin{equation}
\frac{\partial \mathbf{U}}{\partial t} + \frac{\partial \mathbf{F}}{\partial x} = \mathbf{S}
\end{equation}
where the conservative variables $\mathbf{U}$, inviscid fluxes $\mathbf{F}$ and the source terms $\mathbf{S}$ can be written as: 
\begin{align}
\mathbf{U} &= [\rho_g \quad \rho u \quad \rho_g \tilde{e}_g \quad \rho E]^T\\
\mathbf{F} &= [\rho_g u \quad \rho u^2 + p \quad \rho_g u \tilde{e}_g \quad \rho u H]^T\\
\mathbf{S} &= [\omega_g \quad 0 \quad \Omega_g \quad 0]^T
\end{align}
The governing equations are discretized in space using the finite volume method, with inviscid fluxes evaluated through van Leer's flux splitting. A constant reconstruction procedure is applied within each cell, and Strang splitting \cite{strang1968construction} is used for time integration. In this method, the transport operator $\mathcal{T}(\mathbf{U}) = \partial \mathbf{F}/\partial x$ and the reactive operator $\mathcal{R}(\mathbf{U}) = \mathbf{S}$ are integrated sequentially as follows:
\begin{align}
     \frac{\partial \mathbf{U}^{(1)}}{\partial t} &= \mathcal{T}(\mathbf{U}^{(1)}) \enspace, \quad \mathbf{U}^{(1)}(t_n) = \mathbf{U}_n \enspace \label{eq:transp_1}\\
     \frac{\partial \mathbf{U}^{(2)}}{\partial t} &= \mathcal{R}(\mathbf{U}^{(2)}) \enspace, \quad \mathbf{U}^{(2)}(t_n) = \mathbf{U}^{(1)}(t_n + \Delta t/2) \label{eq:react_1}\\
     \frac{\partial \mathbf{U}^{(3)}}{\partial t} &= \mathcal{T}(\mathbf{U}^{(3)}) \enspace, \quad \mathbf{U}^{(3)}(t_n + \Delta t/2) = \mathbf{U}^{(2)}(t_n + \Delta t) \label{eq:transp_2}\\
     \mathbf{U}_{n+1} &= \mathbf{U}^{(3)}(t_n + \Delta t)
\end{align}
where $\Delta t$ denotes the time step. Equation (\ref{eq:react_1}) describes the adiabatic system outlined earlier. The CGM introduces model error and parametric uncertainty in solving this equation. 
Therefore, our goal is to train a surrogate model that learns the solution operator to this adiabatic system, accounting for model and parametric uncertainty, and to replace this equation with the trained surrogate. 

\section{Framework for model predictions with quantified uncertainty}\label{sec:Uncertainty quantification}

The following section outlines the framework used to quantify model and parametric uncertainties in a general setting. Additionally, it describes a surrogate modeling strategy that integrates the Karhunen-Lo\`{e}ve Expansion (KLE) for capturing time dynamics and dimensionality reduction with Polynomial Chaos Expansion (PCE) to map input parameters to the reduced space. This framework is then applied to the CGM setting, where the energy partition parameter accounts for model error and multiplicative factors for grouped rate coefficients address parametric uncertainty. The surrogate model is employed in both inference and uncertainty propagation for 0-D adiabatic reactor and during time-accurate 1-D shock simulations, where it replaces the ODE solver for chemistry.

\subsection{Model error and parametric uncertainty characterization}

Building on the framework of \cite{sargsyan2015statistical}, consider a truth model, or high-fidelity model, denoted by $\boldsymbol{d} = \mathcal{M}^{H}\left(\boldsymbol{\phi}, \boldsymbol{\kappa}_H, t\right)$, where $\boldsymbol{d} \in \mathbb{R}^n$, which best describes the physical process under consideration. Here, $\boldsymbol{\phi}$ represents the set of initial conditions or scenario parameters, $\boldsymbol{\kappa}_H$ represents uncertain parameters (such as rate coefficients in a chemistry model) modeled using a Probability Density Function (PDF), and $t$ denotes an independent variable, such as time. Propagating uncertainty in $\boldsymbol{\kappa}_H$ through the model yields the PDF of quantities of interest, from which moments can be computed.

In general, the high-fidelity model is highly complex and computationally expensive. Therefore, we may consider using a low-fidelity model to describe the physical process, accepting some loss of accuracy in exchange for significantly lower computational costs. To inform the low-fidelity model, a dataset can be generated by sampling from the high-fidelity model, as follows:
\begin{align*}
    d_k\left(\boldsymbol{\phi}_i, t_j\right) \sim \mathcal{N}\left(\mu_k\left(\boldsymbol{\phi}_i, t_j\right), \sigma_k\left(\boldsymbol{\phi}_i, t_j\right)\right), \enspace k = 1, \dots, n 
\end{align*}
where $\mu_k\left(\boldsymbol{\phi}_i, t\right)$ and $\sigma_k\left(\boldsymbol{\phi}_i, t\right)$ are the mean and standard deviation of the $k^{th}$ solution component at time $t_j$, resulting from the uncertainty propagation of $\boldsymbol{\kappa}_H$ through the high-fidelity model for the $i^{th}$ scenario. 

Let the low-fidelity model be denoted by $\boldsymbol{g} = \mathcal{M}^{L}\left(\boldsymbol{\phi}, \boldsymbol{\delta}, \boldsymbol{\kappa}_L, t\right)$ where $\boldsymbol{g} \in \mathbb{R}^n$. Here, $\boldsymbol{\delta} = \{\delta_1, \dots, \delta_d\}$ represents the model reduction parameters that facilitate the construction of the low-fidelity model from the high-fidelity model (and hence, a source of model error). The parameters $\boldsymbol{\kappa}_L$ correspond to the uncertain parameters $\boldsymbol{\kappa}_H$ in the high-fidelity model.
However, because these parameters may be modified during the model reduction process (for example, grouped rate coefficients derived from state-specific rate coefficients), they are distinguished from $\boldsymbol{\kappa}_H$. 

To embed corrections for model error in the low-fidelity model, the model reduction parameters are expanded using a PCE:
\begin{align*}
    \delta_1 &= a_{10} + a_{11}\xi_1\\
    \delta_2 &= a_{20} + a_{21}\xi_1 + a_{22}\xi_2\\
    &\vdots\\
    \delta_d &= a_{d0} + a_{d1}\xi_1 + \dots + a_{dd}\xi_d
\end{align*}
In this work, we employ Hermite-Gauss polynomial-germ pair, so in the above equations, $\boldsymbol{\xi} = \{\xi_i\}_{i=1}^{d}$ represents i.i.d. standard normal variables. This parametrization of $\boldsymbol{\delta}$ can be compactly expressed as $\boldsymbol{\delta} = \boldsymbol{\delta}\left(\boldsymbol{\xi}; \boldsymbol{\alpha}\right)$, where $\boldsymbol{\alpha} = \{a_{jk}\}_{j=1, \dots, d}^{k=0, \dots, j}$ contains the list of PCE coefficients. 
The triangular form of the PCE is chosen to avoid rotational invariance. Additionally, positive priors for the coefficients of $\delta_d$ are selected to prevent multi-modal posteriors, which can arise from simultaneous sign flips of parameters $\{a_{kk}, a_{k+1,k}, \dots, a_{d,k}\}$ \cite{sargsyan2015statistical,sargsyan2019embedded}. The coefficients $\boldsymbol{\alpha}$ can now be inferred along with $\boldsymbol{\kappa}_L$ using data from the high-fidelity model. 
It is important to note that, in the limit of an infinite amount of data, the posterior distribution of the coefficients $\boldsymbol{\alpha}$ (rather than the distributions of $\boldsymbol{\delta}$ themselves) will approach a delta function. This outcome aligns with the fact that predictions from the low-fidelity model will still exhibit model error, even when infinite data is used to infer its parameters.

\subsection{Calibration of model quantities}
The set of model parameters $\boldsymbol{\theta} = \{\boldsymbol{\alpha}, \boldsymbol{\kappa}_L\}$ can be inferred using Bayesian inference. In this framework, current state of knowledge about the system is expressed as a joint prior $p\left(\boldsymbol{\theta} | \mathcal{M}^{L}\right)$. 
After obtaining a dataset $D$ corresponding to several different scenarios from the high-fidelity model, Bayes' theorem is applied to derive the posterior distribution $p\left(\boldsymbol{\theta} | D, \mathcal{M}^{L}\right)$ as follows:
\begin{equation}
    p\left(\boldsymbol{\theta} | D, \mathcal{M}^{L}\right) = \frac{p\left(D | \boldsymbol{\theta}, \mathcal{M}^{L}\right) p\left(\boldsymbol{\theta} | \mathcal{M}^{L}\right)}{p\left(D | \mathcal{M}^{L}\right)}
\end{equation}
where $p\left(D | \boldsymbol{\theta}, \mathcal{M}^{L}\right)$ denotes the likelihood function, which gives the probability of observing the data based on the model $\mathcal{M}^{L}$ and the chosen parameters $\boldsymbol{\theta}$.
The term $p\left(D | \mathcal{M}^{L}\right)$ in the denominator is a normalizing constant, representing the marginalization of the likelihood over the prior, also known as the model evidence, ensuring that the posterior is a proper joint multivariate distribution.
Since multi-dimensional numerical integration is computationally expensive for obtaining posterior PDFs, Markov chain Monte Carlo (MCMC) sampling methods are often used to obtain samples from the posterior distribution \cite{brooks1998markov}.  It is important to note that each sample of $\boldsymbol{\theta}$ results in a probabilistic prediction for a given scenario and temporal location, which is associated with the forward propagation of samples from $\boldsymbol{\xi}$. The mean and variance for the $k^{th}$ QoI can then be computed as:
\begin{align*}
\mu_k\left(\boldsymbol{\phi}, \boldsymbol{\theta}, t\right) = E_{\boldsymbol{\xi}}\left[g_k\left(\boldsymbol{\phi}, \boldsymbol{\delta}\left(\boldsymbol{\xi}; \boldsymbol{\alpha}\right), \boldsymbol{\kappa}_L, t\right)\right]\\
\sigma_k^2\left(\boldsymbol{\phi}, \boldsymbol{\theta}, t\right) = V_{\boldsymbol{\xi}}\left[g_k\left(\boldsymbol{\phi}, \boldsymbol{\delta}\left(\boldsymbol{\xi}; \boldsymbol{\alpha}\right), \boldsymbol{\kappa}_L, t\right)\right]
\end{align*}

Following the approach outlined in \cite{sargsyan2015statistical, sargsyan2019embedded, beaumont2002approximate}, Approximate Bayesian Computation (ABC) is employed to construct a suitable likelihood. The use of full likelihoods or their marginalized approximations often presents challenges, such as degeneracy and multiple posterior singularities. ABC helps mitigate these issues, as detailed in \cite{sargsyan2015statistical, sargsyan2019embedded}. Recently, measure transport has also been proposed for use in Bayesian model calibration when likelihoods are intractable \cite{baptista2024bayesian,marzouk2016introduction}. The applicability of this method for model error calibration will be explored in the future. The likelihood function can be expressed as:
\begin{dmath}
    p\left(D \vert \boldsymbol{\theta}, \mathcal{M}^{L}\right) = \prod_{k=1}^{n} \frac{1}{\varepsilon \sqrt{2 \pi}} \exp \left(-\frac{1}{2 \varepsilon^2}\sum_{i=1}^{N_s}\sum_{j=1}^{N_t} \left(\mu_k\left(\boldsymbol{\phi}_i, \boldsymbol{\theta}, t_j\right) - d_k\left(\boldsymbol{\phi}_i, t_j\right)\right)^2 + \left(\sigma_k\left(\boldsymbol{\phi}_i, \boldsymbol{\theta}, t_j\right) - \gamma\left|\mu_k\left(\boldsymbol{\phi}_i, \boldsymbol{\theta}, t_j\right) - d_k\left(\boldsymbol{\phi}_i, t_j\right)\right| \right)^2 \right) 
\end{dmath}
In the equation above, data is collected at $N_s$ scenario parameters, with each scenario containing $N_t$ points in the temporal domain. This likelihood function favors parameter values for which the model mean closely matches the data, while the model's standard deviation accounts for the discrepancy between the model mean and the observed data. The tolerance parameter $\varepsilon$ determines the extent to which mismatches between the data and model statistics are penalized, and the parameter $\gamma$ scales the model's standard deviation relative to the discrepancy between the model mean and the data. In this study, we set $\varepsilon = 0.5$ and $\gamma = 1$. 

After obtaining the joint posterior distribution for $\boldsymbol{\theta}$, the predictive posterior distributions corresponding to new scenario parameters and temporal locations can be obtained by integrating the model over this posterior distribution: 
\begin{equation}
    p\left(\boldsymbol{g} \vert D, \mathcal{M}^{L}\right) = \int_{\boldsymbol{\theta}} p\left(\boldsymbol{g} \vert \boldsymbol{\theta}, \mathcal{M}^{L}\right)p\left(\boldsymbol{\theta}|D,\mathcal{M}^{L}\right) d\boldsymbol{\theta}
\end{equation}

\subsection{Surrogate modelling for inference and predictions}

We propose a surrogate model that decouples the time dynamics from the influence of other input parameters. Let $g$ represent any component of $\boldsymbol{g} \in \mathbb{R}^n$, where $g \in \{g_1, \dots, g_n\}$. The surrogate model for this QoI can be expressed as: 
\begin{equation}
g\left(\boldsymbol{\phi}, \boldsymbol{\delta}, \boldsymbol{\kappa}_L, t\right) = \sum_{i=1}^{N} c_{i} \psi_{i}\left(\boldsymbol{\phi}, \boldsymbol{\delta}, \boldsymbol{\kappa}_L\right)\eta_{i}\left(t\right)
\end{equation}
Here, $c$ denotes a constant coefficient, $\psi$ and $\eta$ represent normalized modes in the space of input parameters and time respectively, and $N$ denotes the total number of terms in the expansion. 
To achieve this, we utilize the Karhunen-Lo\`{e}ve Expansion (KLE) for the time modes and Polynomial Chaos Expansion (PCE) for the other parameters.
The combination of KLE and PCE has been explored in previous studies \cite{hawchar2017principal, mai2016surrogate, mai2017surrogate} to address time-variant reliability problems and nonlinear oscillatory systems. However, to the best of the author's knowledge, this method has not been tested for operator learning.

\subsubsection{Karhunen-Lo\`{e}ve Expansion}

For a stochastic process $g(\boldsymbol{\phi}, \boldsymbol{\delta}, \boldsymbol{\kappa}_L, t)$ defined on the probability space $(\Omega, \mathcal{F}, P)$ - where $\Omega$ is the sample space, $\mathcal{F}$ is the $\sigma$-algebra of events and $P: \mathcal{F} \rightarrow [0,1]$ is a probability measure, the Karhunen-Lo\`{e}ve expansion can be expressed as 
\begin{equation*}
g(\boldsymbol{\phi}, \boldsymbol{\delta}, \boldsymbol{\kappa}_L, t) = \overline{g}(t) + \sum_{i=1}^{\infty} \sqrt{\lambda_i} \nu_i(\boldsymbol{\phi}, \boldsymbol{\delta}, \boldsymbol{\kappa}_L)\eta_i(t)
\end{equation*}
where $\overline{g}(t)$ denotes the mean of the stochastic process, $\eta_i$ and $\lambda_i$ are the eigenfunctions and eigenvalues, respectively, of the covariance function of the stochastic process, obtained by solving the eigenvalue problem:
\begin{equation*}
    \int_{\mathcal{D}_t} \mathcal{R}(t, t') \eta_i(t') dt = \lambda_i \eta_i(t)
\end{equation*}
where $\mathcal{R}(t, t')$ represents the covariance function and $\mathcal{D}_t$ denotes the temporal domain. 
The eigenmodes $\eta_i$ form the basis in time for the proposed surrogate model. 
The decay rate of eigenvalues is linked to the correlation length of the process being expanded. 
The random coefficients $\nu_i(\boldsymbol{\phi}, \boldsymbol{\delta}, \boldsymbol{\kappa}_L)$ are centered and orthonormal, satisfying:
\begin{align*}
    E\{\nu_i\} &= 0 \quad \forall i\\
    E\{\nu_i \nu_j\} &= \delta_{ij} \quad \forall i, j
\end{align*}
where $\delta$ is the Kronecker delta function. 
These coefficients, generally not independent, are obtained as:
\begin{equation*}
\nu_i(\boldsymbol{\phi}, \boldsymbol{\delta}, \boldsymbol{\kappa}_L) = \frac{1}{\sqrt{\lambda_i}} \int_{\mathcal{D}_t} \left[g(\boldsymbol{\phi}, \boldsymbol{\delta}, \boldsymbol{\kappa}_L, t) - \overline{g}(t)\right] \eta_i(t) dt
\end{equation*}
KLE is guaranteed to converge in a mean-squared sense for any process with finite variance \cite{loeve1977elementary}. Given that the quantities of interest stem from solving a physical system, it is expected that a truncated series with few terms can accurately represent the stochastic process:
\begin{equation}
g(\boldsymbol{\phi}, \boldsymbol{\delta}, \boldsymbol{\kappa}_L, t) = \overline{g}(t) + \sum_{i=1}^{P} \sqrt{\lambda_i} \nu_i(\boldsymbol{\phi}, \boldsymbol{\delta}, \boldsymbol{\kappa}_L)\eta_i(t)
\end{equation} 
To make predictions for a test set of input parameters, a mapping from parameter space $\{\boldsymbol{\phi}, \boldsymbol{\delta}, \boldsymbol{\kappa}_L\}$ to the random variables $\nu_i$ is required. This is achieved using PCE, as detailed in the next section. 

\subsubsection{Polynomial Chaos Expansion}

The Wiener-Hermite chaos \cite{wiener1938homogeneous} was the first representation of a stochastic process as an expansion in orthogonal polynomials. While this expansion converges for any second-order stochastic process, it is optimal only for Gaussian inputs \cite{cameron1947orthogonal}. To address a wider range of stochastic processes, the Wiener-Askey polynomial chaos, or generalized polynomial chaos (gPC), was introduced. In this approach, the PDFs of various standard random inputs are matched with the weight functions of suitable orthogonal polynomials from the Askey scheme, enabling exponential convergence in representing the stochastic process \cite{xiu2002wiener}. The gPC method has been successfully applied to numerous engineering problems, validating its effectiveness \cite{ghanem2003stochastic,najm2009uncertainty,sudret2014polynomial,shen2020polynomial}.
 
Using $\nu$ to denote a KLE coefficient, it is mapped to the random inputs using an orthogonal polynomial expansion:
\begin{equation*}
\nu(\boldsymbol{\phi}, \boldsymbol{\delta}, \boldsymbol{\kappa}_L) = \sum_{\boldsymbol{\alpha} \in \mathcal{J}^{M, p}_{q}} a_{\boldsymbol{\alpha}} \psi_{\boldsymbol{\alpha}}(\boldsymbol{\xi})
\end{equation*}
Here, 
the random inputs $\{\boldsymbol{\phi}, \boldsymbol{\delta}, \boldsymbol{\kappa}_L\}$ are first mapped component-wise to a vector of standard normal random variables $\boldsymbol{\xi}$ using the standard one-dimensional inverse CDF method. The multi-index $\boldsymbol{\alpha}$ belongs to a set of indices obtained using an appropriate truncation strategy. In this work, we use the hyperbolic index set $\mathcal{J}^{M, p}_{q} \equiv \{\boldsymbol{\alpha} \in \mathbb{N}^M : ||\boldsymbol{\alpha}||_q \leq p\}$, where $M$ is the input dimensionality; $q$ (with $0 < q < 1$) penalizes the high-rank indices based on the $q$-norm, and $p$ is the maximum polynomial degree.
In this work, we set $q = 0.8$ for all quantities of interest.  
The basis functions $\psi_{\boldsymbol{\alpha}}$ are normalized multi-variate Hermite polynomials, obtained from tensor products of uni-variate polynomials:
\begin{equation*}
\psi_{\boldsymbol{\alpha}}(\boldsymbol{\xi}) = \prod_{i=1}^{M} \psi_{\alpha_i} (\xi_i)
\end{equation*}
where 
\begin{equation*}
    \psi_{n}(\xi) = \frac{h_{n}(\xi)}{\sqrt{n!}}
\end{equation*}
Here, $h_n$ is the standard one-dimensional Hermite polynomial of order $n$ with norm $||h_n|| = \sqrt{n!}$, $n \in \mathbb{N}$. The normalized multi-variate Hermite polynomials are orthonormal with respect to the joint PDF of input parameters $p(\boldsymbol{\xi}) = \prod_{i=1}^{M} p(\xi_i)$:
\begin{equation*}
    \langle \psi_{\boldsymbol{\alpha}}, \psi_{\boldsymbol{\beta}} \rangle = \int_{\mathbb{R}^M} \psi_{\boldsymbol{\alpha}}(\boldsymbol{\xi}) \psi_{\boldsymbol{\beta}}(\boldsymbol{\xi}) p(\boldsymbol{\xi})d\boldsymbol{\xi} = \delta_{\boldsymbol{\alpha}, \boldsymbol{\beta}} 
\end{equation*}
where $\delta$ is the Kronecker delta function. 
The PCE coefficients $a_{\boldsymbol{\alpha}}$ can be computed using either the spectral projection method \cite{smith2013uncertainty}, which exploits the orthogonality of truncated basis and uses quadrature approximations or by the regression method. 

\subsubsection{Random PCE coefficients with Extended PCE formulation}\label{sec:rand_pce_coef}

The extended PCE formulation \cite{wang2023stochastic,wang2021extended,wang2022functional} allows us to separate the effect of model error parameters from aleatoric uncertainty parameters:
\begin{align*}
\nu(\boldsymbol{\phi}, \boldsymbol{\delta}, \boldsymbol{\kappa}_L) = \sum_{\boldsymbol{\alpha} \in \mathcal{J}^{M, p}_{q}} a_{\boldsymbol{\alpha}_{\sim \boldsymbol{\delta}}\boldsymbol{\alpha}_{\boldsymbol{\delta}}} \psi_{\boldsymbol{\alpha}_{\sim \boldsymbol{\delta}}}(\boldsymbol{\xi}_{\sim \boldsymbol{\delta}}) \psi_{\boldsymbol{\alpha}_{\boldsymbol{\delta}}}(\boldsymbol{\xi}_{\boldsymbol{\delta}})
\end{align*}
Here $|\boldsymbol{\alpha}| = |\boldsymbol{\alpha}_{\sim \boldsymbol{\delta}}| + |\boldsymbol{\alpha}_{\boldsymbol{\delta}}|$; $\psi_{\boldsymbol{\alpha}_{\sim \boldsymbol{\delta}}}(\boldsymbol{\xi}_{\sim \boldsymbol{\delta}})$ and $\psi_{\boldsymbol{\alpha}_{\boldsymbol{\delta}}}(\boldsymbol{\xi}_{\boldsymbol{\delta}})$ represent terms corresponding to aleatoric and model error (epistemic) parameters respectively. 
This can be further represented as a PCE with coefficients dependent on the model error parameter as: 
\begin{align*}
\nu(\boldsymbol{\phi}, \boldsymbol{\delta}, \boldsymbol{\kappa}_L) = \sum_{\boldsymbol{\alpha} \in \mathcal{J}^{M, p}_{q}} a_{\boldsymbol{\alpha}_{\sim \boldsymbol{\delta}}}(\boldsymbol{\xi}_{\boldsymbol{\delta}}) \psi_{\boldsymbol{\alpha}_{\sim \boldsymbol{\delta}}}(\boldsymbol{\xi}_{\sim \boldsymbol{\delta}})
\end{align*}
By separating the ${\boldsymbol{\delta}}$-dependent and independent parts, we get:
\begin{align*}
\nu(\boldsymbol{\phi}, \boldsymbol{\delta}, \boldsymbol{\kappa}_L) = \underbrace{\sum_{\substack{\boldsymbol{\alpha} \in \mathcal{J}^{M, p}_{q} \\ |\boldsymbol{\alpha}_{\boldsymbol{\delta}}| = 0}} a_{\boldsymbol{\alpha}_{\sim \boldsymbol{\delta}}} \psi_{\boldsymbol{\alpha}_{\sim \boldsymbol{\delta}}}(\boldsymbol{\xi}_{\sim \boldsymbol{\delta}})}_{\boldsymbol{\delta} \text{-independent part}} + \underbrace{\sum_{\substack{\boldsymbol{\alpha} \in \mathcal{J}^{M, p}_{q} \\ |\boldsymbol{\alpha}_{\boldsymbol{\delta}}| \neq 0}} a_{\boldsymbol{\alpha}_{\sim \boldsymbol{\delta}}} (\boldsymbol{\xi}_{\boldsymbol{\delta}}) \psi_{\boldsymbol{\alpha}_{\sim \boldsymbol{\delta}}}(\boldsymbol{\xi}_{\sim \boldsymbol{\delta}})}_{\boldsymbol{\delta} \text{-dependent part}}
\end{align*}
The $\boldsymbol{\delta}$-dependent component of the surrogate addresses the model error of the low-fidelity model. Furthermore, this model error is incorporated in a physically consistent manner. Specifically, the $\boldsymbol{\delta}$-dependent part remains influenced by the initial conditions $\boldsymbol{\phi}$ through the germ $\boldsymbol{\xi}_{\sim \boldsymbol{\delta}}$. This indicates that the model error is not treated as a constant term, as in some explicit model error methodologies, but rather as a function of the scenario parameters. 

\begin{figure}[H]
    \centering
    \subcaptionbox{Training phase}{\includegraphics[width=0.42\linewidth]{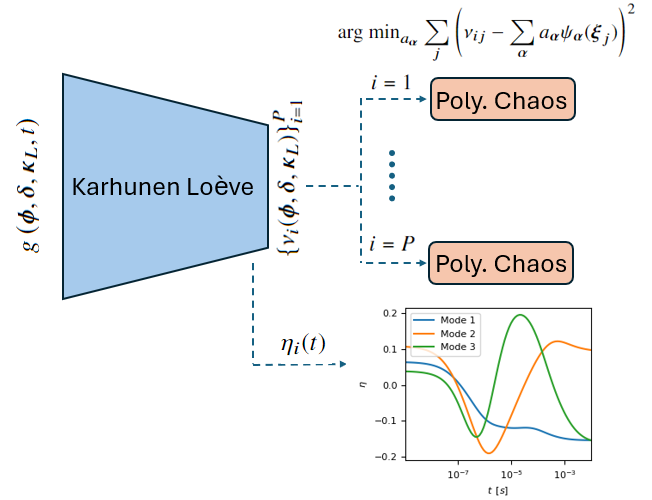}}
    \subcaptionbox{Prediction phase}  {\includegraphics[width=0.48\linewidth]{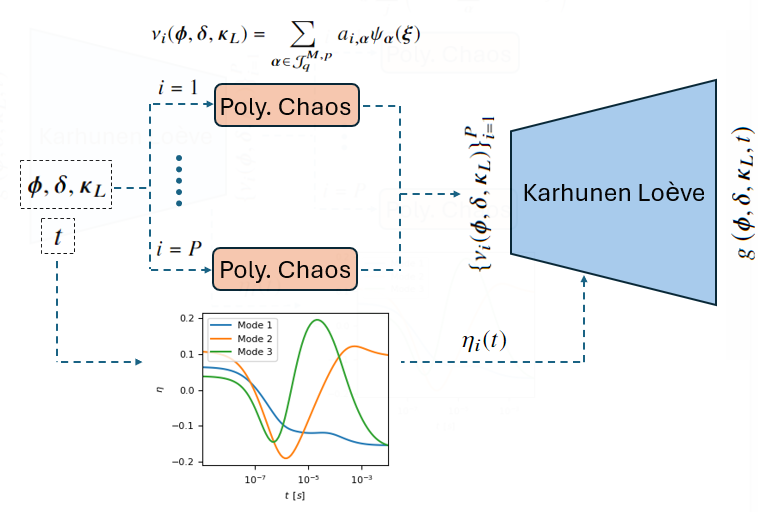}}
    \caption{Surrogate modelling methodology}
    \label{fig:surrogate_modelling}
\end{figure}

The overall surrogate model is obtained by combining the KLE and PCE components as follows:
\begin{equation}
g\left(\boldsymbol{\phi}, \boldsymbol{\delta}, \boldsymbol{\kappa}_L, t\right) = \overline{g}(t) + \sum_{i=1}^{P} \sqrt{\lambda_i} \left[\sum_{\boldsymbol{\alpha} \in \mathcal{J}^{M, p}_{q,i}} a_{i,\boldsymbol{\alpha}} \psi_{\boldsymbol{\alpha}}(\boldsymbol{\xi})\right]\eta_i(t)
\label{eq:overall_surrogate}
\end{equation}
Here, the subscript $i$ distinguishes the multi-index set $\mathcal{J}^{M, p}_{q,i}$ and PCE coefficients $a_{i,\boldsymbol{\alpha}}$ for each KLE coefficient, indicating that each KLE coefficient has its own PCE expansion. Separate surrogates of this form are constructed for each QoI in the system, which are interrelated only through common inputs. An illustration for the surrogate modeling methodology is shown in Figure \ref{fig:surrogate_modelling}.    

\subsection{Application to non-equilibrium chemical kinetics}

The framework developed is applied to the CGM for non-equilibrium chemical kinetics of the $\mathrm{O}_2$-$\mathrm{O}$ system. The low-fidelity model ($\mathcal{M}^L$) is the 2-bin energy-based model, while the high-fidelity model ($\mathcal{M}^H$) is the 10-bin spectral clustering model. The high-fidelity model provides data to capture model error and parametric uncertainty in the 2-bin model. Parametric uncertainty, $\boldsymbol{\kappa}_H$, arises from the uncertainty in the rate coefficients for dissociation and excitation reactions. Independent multiplicative factors with a log-uniform distribution between 0.2 and 5 are applied to the nominal rate coefficients when generating the dataset.

We focus on the model error introduced by the grouping strategy. Specifically, the energy partition parameter, $\delta$, which divides the groups in the 2-bin model, is treated as the model error parameter. Introducing stochasticity into $\delta$ and expanding it in a first-order PCE, we have: 
\begin{align}\label{eq:eng_part_pce}
    \delta = a_{10} + a_{11}\xi
\end{align}
where $\xi$ is a standard normal random variable. The expansion coefficients $a_{10}$ and $a_{11}$ are the inference targets.
Rate coefficient multiplicative factors for grouped dissociation and excitation reactions are also calibrated to account for parametric uncertainty. Given the dependency of grouped rate coefficients for mass and energy, the same multiplicative factor is used for grouped mass and energy rate coefficients derived from the same state-specific rate coefficients. This results in one multiplicative factor for each dissociation and excitation reaction: $\boldsymbol{\kappa}_L = \{k_{d1}, k_{d2}, k_{e12}\}$. The reverse rate coefficients are derived from micro-reversibility.
The calibration parameters are thus $\boldsymbol{\theta} = \{a_{10}, a_{11}, k_{d1}, k_{d2}, k_{e12}\}$ and their prior distributions are listed in Table \ref{tab:table_priors}.

Rovibrational state-specific population evolution is the most accurate choice to learn model error and parametric uncertainty in reduced order models as this quantify captures all aspects of thermochemical non-equilibrium. However, due to its high dimensionality (over 6000 states for $\mathrm{O}_2$), we use mass fractions in six uniformly spaced internal energy bins, $Y^{sts}_j|_{j=1}^6$, as the quantities of interest to inform the low-fidelity model. The quantities of interest for prediction in CFD, as described in \ref{sec:shock_goveq}, employing the 2-bin model include two group species mass fractions, two group internal temperatures, and the translational temperature. Hence, the set of observables can be represented by $\boldsymbol{g} = \{Y^{sts}_1, Y^{sts}_2, Y^{sts}_3, Y^{sts}_4, Y^{sts}_5, Y^{sts}_6, Y_1, Y_2, T, T_1, T_2\}$, where we reiterate that the first six are used to learn the model error and parametric uncertainty and the remaining observables are for prediction in CFD simulations. The set of initial conditions includes density, translational temperature, internal temperatures, and species mass fractions at the previous time step, giving $\boldsymbol{\phi} = \{\rho, T, T_1, T_2, Y_1, Y_2\}$.   

To expedite the inference process and for predictions in CFD with quantified uncertainties, we construct a surrogate model for each QoI following the methodology described previously.
We note that the surrogates are constructed using the prior distributions of model and parametric uncertainty, however during predictions in CFD, the posterior distributions of model and parametric uncertainty is forward propagated through the surrogate.  
The training and test data generation procedure is described in the next section. 

\begin{center}
\captionof{table}{Priors for calibration parameters}
\label{tab:table_priors}
\begin{tabular}{|l|l|l|l|l|l|}
\hline
Parameter        &  $a_{10}\, [eV]$ &
$a_{11}\, [eV]$ &$k_{d1}\, [-]$ & $k_{d2}\, [-]$ & $k_{e12}\, [-]$\\
\hline
Distribution    & U[2, 5] & 
U[0.3, 2] &
logU[0.2, 5] & logU[0.2, 5] & logU[0.2, 5] \\
\hline
\end{tabular}
\end{center}

\begin{center}
\captionof{table}{Range of initial conditions}
\label{tab:table_range_ics}
\begin{tabular}{|l|l|l|l|l|l|l|}
\hline
Parameter        &  $\rho\, [kg/m^{3}]$ &  $T\, [K]$ & $T_{1}\,[K]$ & $T_{2}\, [K]$ & $Y_1\,[-]$ & $Y_2\,[-]$\\
\hline
min    & 0.01 & 300 & 300 & 300 & 0.05 & $10^{-8}$\\
\hline
max    & 0.20 & 12000 & 8000 & 8000 & 1.0 & 0.2\\
\hline
\end{tabular}
\end{center}

\subsection{Training and test data generation}
As previously stated, we aim to use the surrogate model for a range of initial conditions in 0-D adiabatic reactor and time accurate 1-D shock simulations. To achieve this, we generate 100 samples for free-stream pressure ranging from [500 to 1000] $Pa$, temperature fixed to 300 K, and velocity ranging from [4 to 7] km/s using LHS, in a 1-D shock scenario. For each free-stream sample, we conduct time-accurate shock simulations using samples from the joint prior of the model error parameter $\delta$ and the multiplicative factors for grouped rate coefficients, $\boldsymbol{\kappa}_L$, applying Strang splitting. 
It is noted that the prior distributions of $a_{10}$ and $a_{11}$ are used to compute the prior distribution of the energy partition parameter $\delta$ via equation (\ref{eq:eng_part_pce}). This distribution is then used to generate samples for building the surrogate model. 
We collect the initial conditions $\boldsymbol{\phi} = \{\rho, T, T_1, T_2, Y_1, Y_2\}$ from the fluid solver to the reactive operator, resulting in a training dataset of approximately 100,000 samples. The resulting range of initial conditions is shown in Table \ref{tab:table_range_ics}. We then perform 0-D adiabatic reactor simulations over the time interval $t = [0, 0.01] \enspace \text{s}$ for these initial conditions to obtain the time evolution of the QoIs, $\boldsymbol{g}(\boldsymbol{\phi}, \delta, \boldsymbol{\kappa}_L, t) \in \{T, T_1, T_2, Y_1, Y_2\}$.
Data points are log-uniformly sampled in the time domain, focusing on regions where excitation and dissociation reactions are most active, while fewer points are taken as equilibrium is approached. This approach ensures data is collected from temporal regions of high variability in the QoIs.
The trained surrogate model is then tested on initial condition samples drawn from Table \ref{tab:table_range_ics} for the 0-D adiabatic reactor and from free-stream conditions for 1-D shock that were not present in the training dataset.    
Lastly, we note that although the random inputs generated in this manner may be correlated, we treat them as independent and use tensor-product measure dominating the true dependent measure in the PCE formulation \cite{jakeman2010numerical,chen2013flexible,jakeman2019polynomial}.

\section{Results and discussions}\label{sec:Results and Discussions}

In this section, we evaluate the performance of the proposed framework for model error characterization and surrogate modeling for reduced-order models of non-equilibrium chemistry. First, we analyze the results from Bayesian inference, accounting for uncertainties in the energy partition parameter and rate coefficients in the 2-bin energy-based model. The resulting posteriors are then propagated through the low-fidelity model to obtain predictive posteriors that capture both model error and parametric uncertainty. Next, we leverage the surrogate model to assess the importance of each input parameter in predicting the solution at the next time step, using analytically derived first and total-order Sobol indices. Finally, we test the surrogate's predictive performance in a 0-D adiabatic reactor and a 1-D shock case, comparing results to those obtained through conventional numerical integration using the second-order Backward Differentiation Formula (BDF2).

In this work, we use the PyMC package \cite{abril2023pymc} with the No-U-Turn sampler (NUTS) \cite{hoffman2014no,neal2012mcmc,betancourt2017conceptual} to sample posterior distributions of inference parameters. The availability of analytical likelihood derivatives from the surrogate model makes NUTS an efficient choice. 
As previously mentioned, a separate surrogate model (KLE followed by PCE for each KLE coefficient) is constructed for each QoI. The number of KLE modes is selected to keep the average relative L2 reconstruction error below $10^{-4}$ (see \ref{sec:Appendix-KLE-truncation}). An adaptive LAR approach \cite{blatman2009adaptive,blatman2009anisotropic} was used to determine the PCE coefficients, with the PCE order adaptively increased up to 10 until a variance-normalized leave-one-out cross-validation error (LOOCV) of \( 10^{-6} \) was achieved, while mitigating overfitting. The PCE coefficients were computed using an in-house developed Stochastic Modelling and Uncertainty Quantification (SMUQ) toolbox \cite{rostkowski2019calibration,rostkowski2022quantification,venturi2014physics}.

\subsection{Bayesian inference}

\begin{figure}[H]
    \centering
    \subcaptionbox{\label{fig:joint_posterior}}{\includegraphics[width=0.59\linewidth]{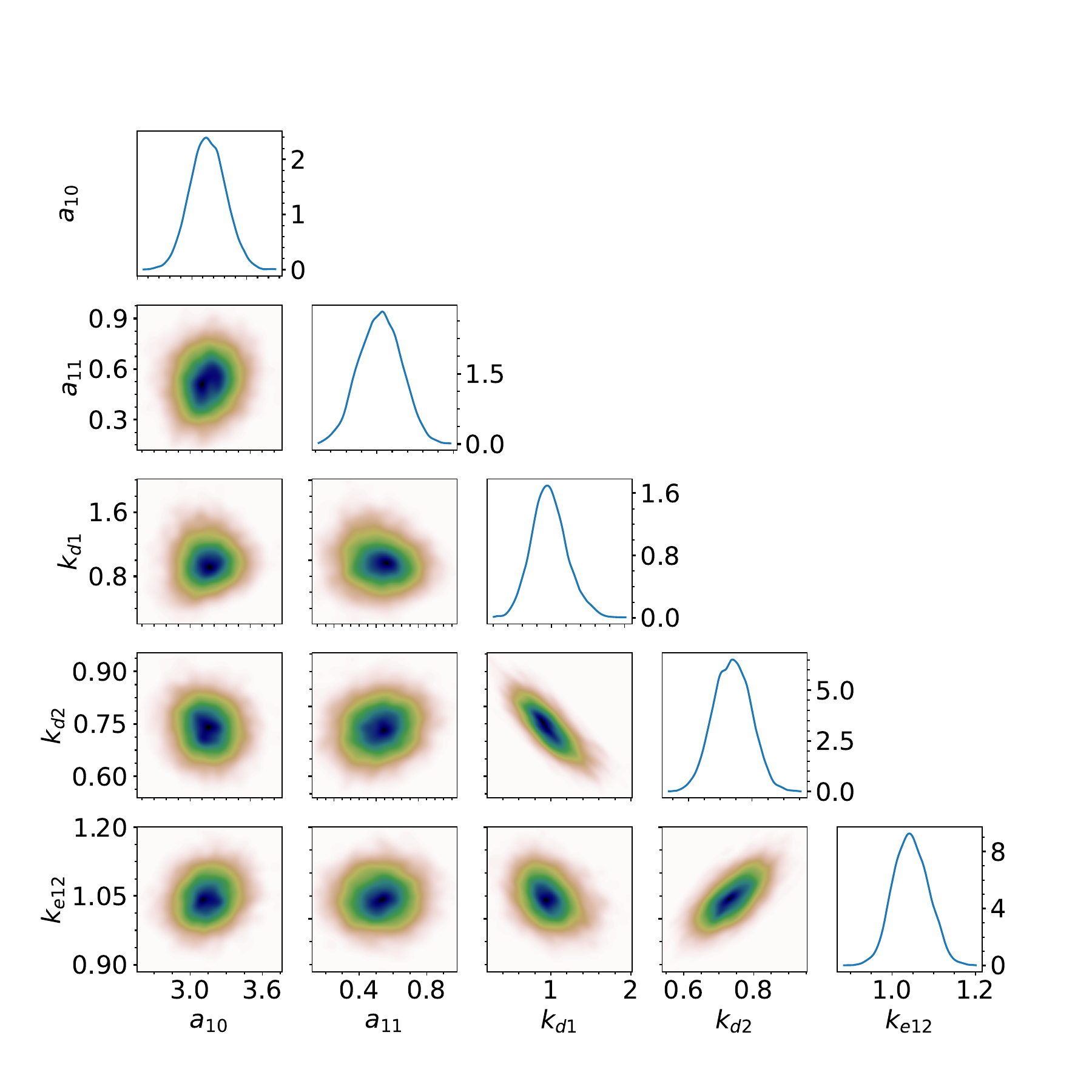}}
    \hspace{0.1cm}
    \subcaptionbox{\label{fig:energy_partition}}{\includegraphics[width=0.39\linewidth]{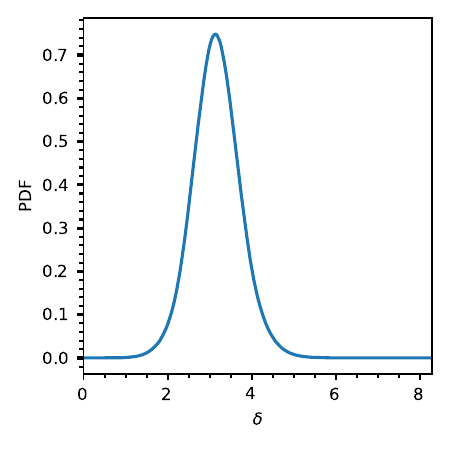}}
    \caption{Bayesian inference results. Joint posterior distribution (a) and energy partition parameter posterior distribution from \cref{eq:eng_part_pce} (b).}
    \label{fig:bayesian_inference}
\end{figure}

Fig. \ref{fig:joint_posterior} shows the joint posterior distribution obtained from two MCMC chains and a total of 20000 chain samples with a burn-in of 2000 samples per chain. We see that the two parameters $a_{10}$ and $a_{11}$ governing the model error have almost no correlation for the chosen scenario parameters. Notably, there is also little correlation between these model error parameters and the rate coefficient multiplicative factors $\boldsymbol{\kappa}_L$, which capture parametric uncertainty. A moderately negative correlation is observed between the dissociation rate coefficients $k_{d1}$ and $k_{d2}$, while a positive correlation is found between $k_{d2}$ and $k_{e12}$. This suggests that, although the rate coefficient parameters are independent in the high-fidelity model, some dependence arises at the low-fidelity level. The marginal posterior distributions of both $k_{d1}$ and $k_{e12}$ support values greater than one, implying faster dissociation from group-1 and faster excitation between groups 1 and 2. In contrast, the marginal posterior of $k_{d2}$ is mostly below one, indicating slower-than-nominal dissociation from group-2. It is important to note that model error in the low-fidelity model influences the posterior distributions of the rate coefficients, so these posterior distributions should not be interpreted as physically meaningful.
Propagating the posteriors of $a_{10}$ and $a_{11}$ through \cref{eq:eng_part_pce} yields the posterior distribution of the energy partition parameter, shown in Fig. \ref{fig:energy_partition}. The MAP value is around 3 eV, which is consistent with the energy threshold for enhanced relaxation processes, where beyond this threshold exchange reactions play a dominant role in the system dynamics. Majority of the support lies below 4 eV, suggesting that a non-uniform energy-based model with a smaller energy interval for the first bin performs better than a uniform model with equal energy intervals.    

\begin{figure}[htbp]
    \centering
    \subcaptionbox{\label{fig:post_pred_Y1}}{\includegraphics[width=0.49\linewidth]{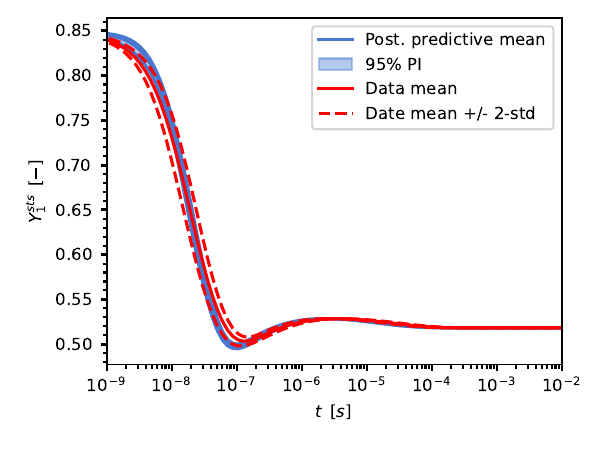}}
    \subcaptionbox{\label{fig:post_pred_Y2}}{\includegraphics[width=0.49\linewidth]{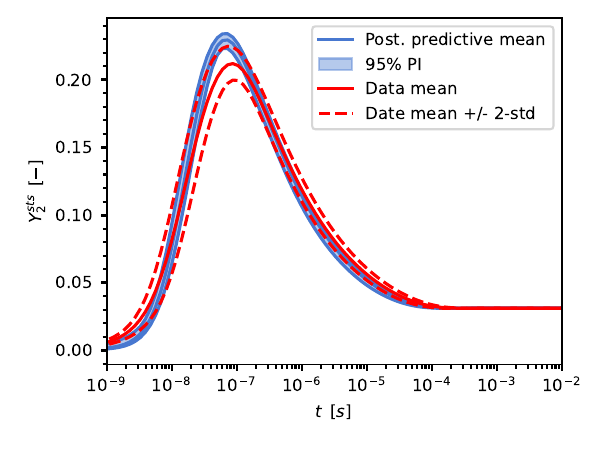}}
    \subcaptionbox{\label{fig:post_pred_Y3}}{\includegraphics[width=0.49\linewidth]{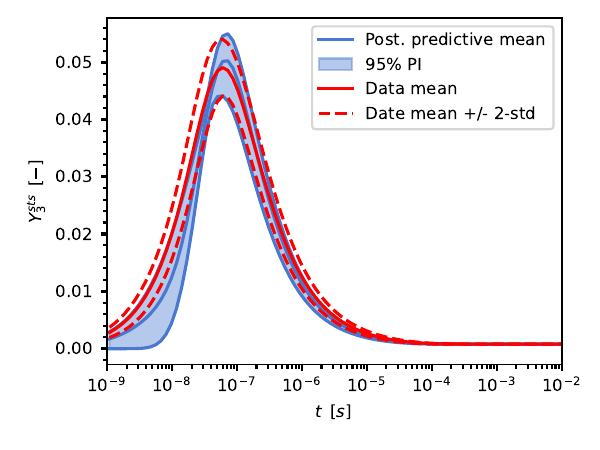}}
    \subcaptionbox{\label{fig:post_pred_Y4}}{\includegraphics[width=0.49\linewidth]{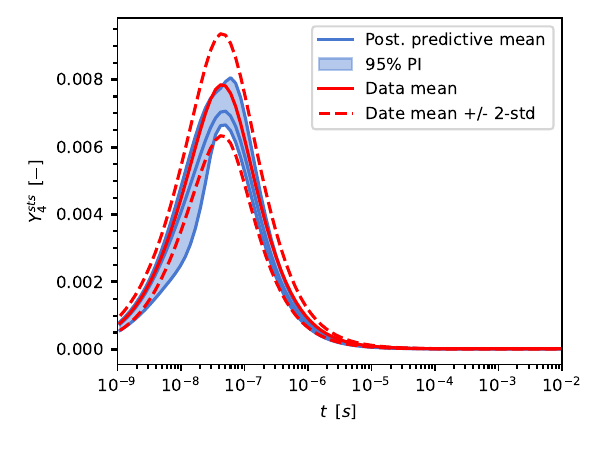}}
    \subcaptionbox{\label{fig:post_pred_Y5}}{\includegraphics[width=0.49\linewidth]{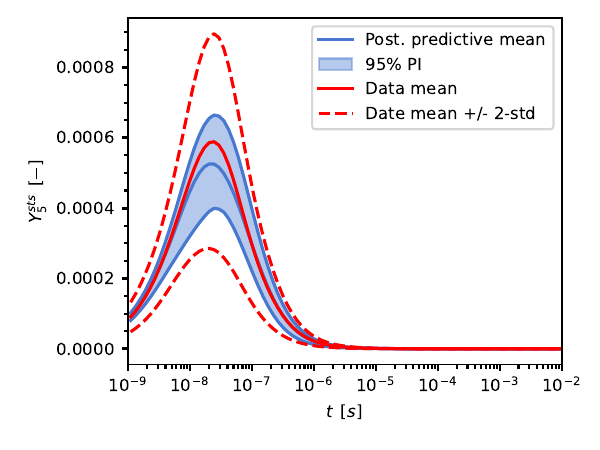}}
    \subcaptionbox{\label{fig:post_pred_Y6}}{\includegraphics[width=0.49\linewidth]{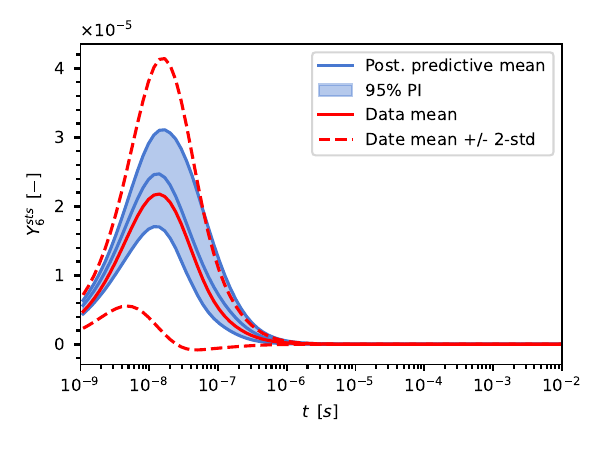}}
    \caption{Predictive posteriors for mass fractions of the six uniform energy bins for initial conditions $\rho = 0.035 \, kg/m^3, T = 11000 \, K, T_1 = T_2 = 1500 \, K, Y_1 = 0.85, Y_2 = {10}^{-8}$.}
    \label{fig:post_pred}
\end{figure}

Fig. \ref{fig:post_pred} shows the posterior distributions for $Y_1^{sts}$ through $Y_6^{sts}$, obtained by propagating the joint posterior (Fig. \ref{fig:joint_posterior}) through the low-fidelity model in a 0-D adiabatic reactor for a given set of initial conditions. We see that the probabilistic predictions from the low-fidelity model are able to capture the high-fidelity solution. We still see some discrepancy between the corrected low-fidelity model and the high-fidelity solution which can be attributed to other sources of model error, such as truncation of reconstruction strategy, which have not been accounted for in this work.    

\subsection{Sensitivity analysis} 
\begin{figure}[H]
    \centering
    \subcaptionbox{$T$}{\includegraphics[width=0.49\linewidth]{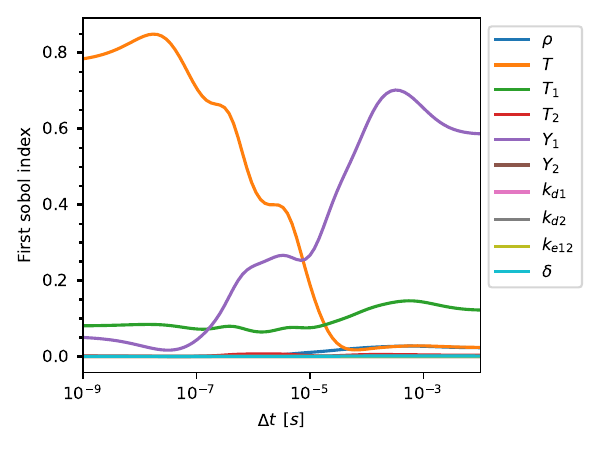}}
    \hspace{0.1cm}
    \subcaptionbox{$T_1$}{\includegraphics[width=0.49\linewidth]{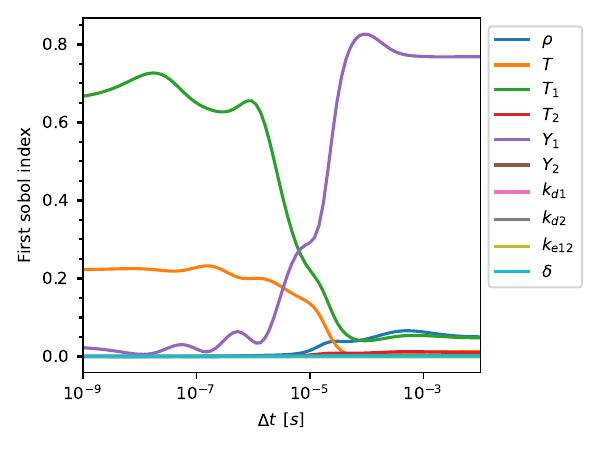}}
    \subcaptionbox{$T_2$}{\includegraphics[width=0.49\linewidth]{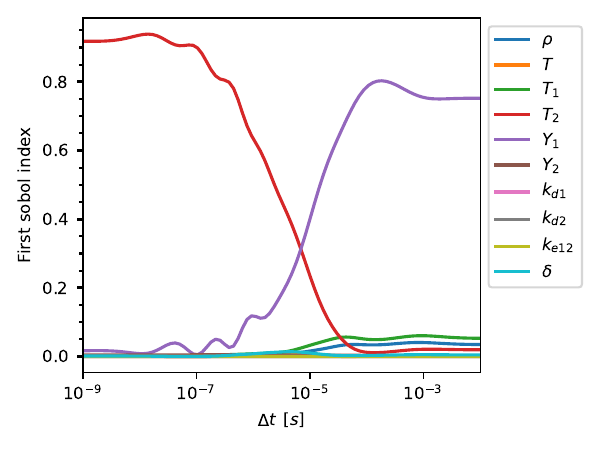}}
    \hspace{0.1cm}
    \subcaptionbox{$Y_1$}{\includegraphics[width=0.49\linewidth]{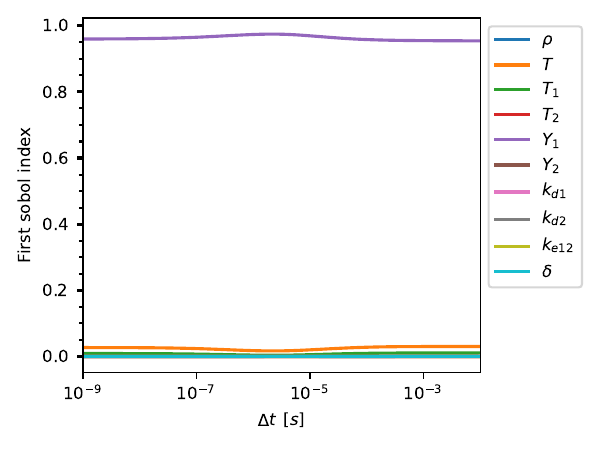}}
    \hspace{0.1cm}
    \subcaptionbox{$Y_2$}{\includegraphics[width=0.49\linewidth]{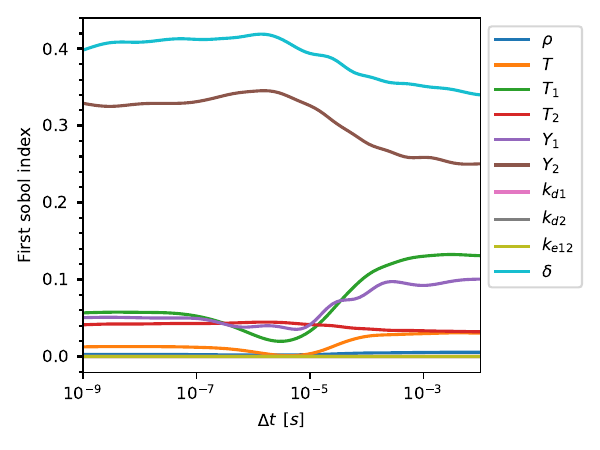}}
    \caption{First-order Sobol indices post-processed from surrogate as a function of time step.}
    \label{fig: first_sobol}
\end{figure}

\begin{figure}[H]
    \centering
    \subcaptionbox{$T$}{\includegraphics[width=0.49\linewidth]{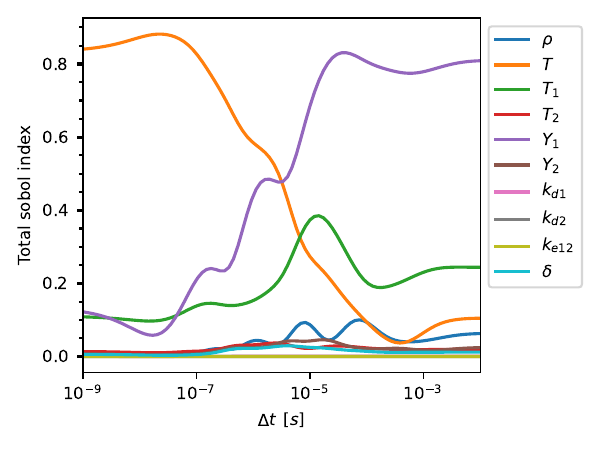}}
    \hspace{0.1cm}
    \subcaptionbox{$T_1$}{\includegraphics[width=0.49\linewidth]{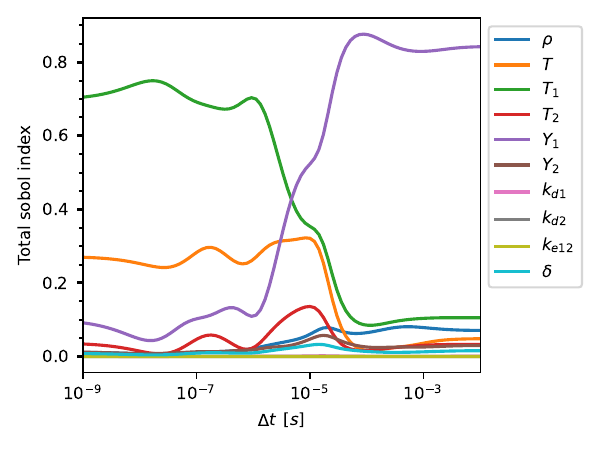}}
    \subcaptionbox{$T_2$}{\includegraphics[width=0.49\linewidth]{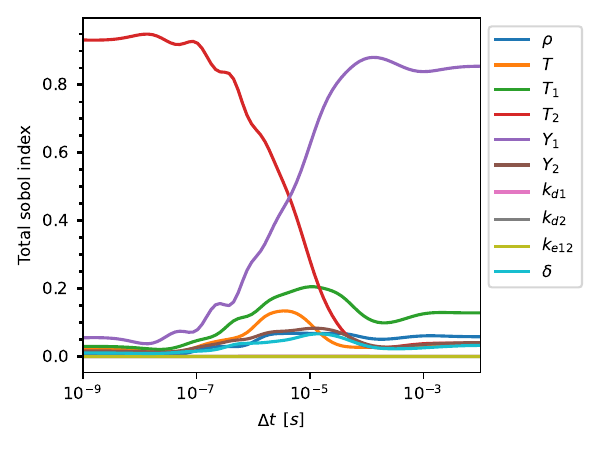}}
    \hspace{0.1cm}
    \subcaptionbox{$Y_1$}{\includegraphics[width=0.49\linewidth]{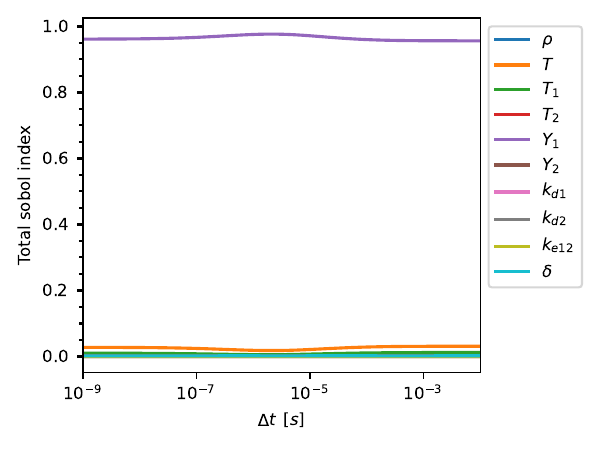}}
    \hspace{0.1cm}
    \subcaptionbox{$Y_2$}{\includegraphics[width=0.49\linewidth]{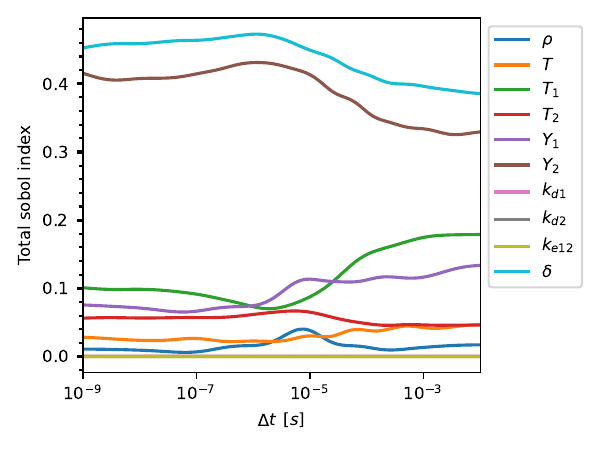}}
    \caption{Total-order Sobol indices post-processed from surrogate as a function of time step.}
    \label{fig: total_sobol}
\end{figure}

PCE allows for the analytical computation of Sobol indices directly from its coefficients. This enables us to rank the influence of input parameters at the previous time step on the quantities of interest at the next time step when using the surrogate model for time marching. To compute Sobol indices over time steps, the KLE modes in time must be multiplied with the PCE coefficients before calculating the Sobol indices, as shown below. We begin with \cref{eq:overall_surrogate}, rewritten here with the KLE modes combined with the PCE coefficients:
\begin{align}
g\left(\boldsymbol{\phi}, \boldsymbol{\delta}, \boldsymbol{\kappa}_L, t\right) &= \overline{g}(t) + \sum_{i=1}^{P} \sum_{\boldsymbol{\alpha} \in \mathcal{J}^{M, p}_{q,i}} \sqrt{\lambda_i} a_{i,\boldsymbol{\alpha}} \eta_i(t) \psi_{\boldsymbol{\alpha}}(\boldsymbol{\xi})
\end{align}
The first and total Sobol indices of the QoI $g$ with respect to parameter $k$ can be computed as a function of the step size $t$ as:
\begin{align}
    S^F_k(t) &= \frac{\sum_{i=1}^P \sum_{\boldsymbol{\alpha} \in \mathcal{I}_k} \left(\sqrt{\lambda_i} a_{i, \boldsymbol{\alpha}} \eta_i(t)\right)^2}{\text{Var}_g(t)} \\
    S^T_k(t) &= \frac{\sum_{i=1}^P \sum_{\boldsymbol{\alpha} \in \mathcal{J}_k} \left(\sqrt{\lambda_i} a_{i, \boldsymbol{\alpha}} \eta_i(t)\right)^2}{\text{Var}_g(t)} \\
    \text{Var}_g(t) &= \sum_{i=1}^P\sum_{\boldsymbol{\alpha} \in \mathcal{J}^{M, p}_{q,i} / \, {\boldsymbol{0}}} \left(\sqrt{\lambda_i} a_{i, \boldsymbol{\alpha}} \eta_i(t)\right)^2
\end{align}
where $\mathcal{I}_k$ denotes the set of indices where only the $k$th entry of the multi-index is non-zero and the remaining entries are zero and $\mathcal{J}_k$ denotes the set of indices where the $k$th entry of the multi-index is non-zero and the remaining entries are either zero or non-zero. $\text{Var}_g$ denotes the variance of $g$ which is again obtained from all the PCE coefficients except the constant term. 
Figures \ref{fig: first_sobol} and \ref{fig: total_sobol} show the first and total order Sobol indices for QoIs $T$ through $Y_2$, that are required for predictions in CFD. While there are quantitative differences between the two indices for all QoIs, the qualitative agreement suggests that interaction effects between the parameters have a minimal impact on the variability of the QoIs. 
The total order Sobol index plots for $T$, $T_1$ and $T_2$ show that at small time steps, the variability in their initial value is the most influential factor. However, at larger time steps, the mass fraction of group-1 becomes more influential. 
Clearly, smaller time steps indicate that the system starting from its initial state, which suppose is out of equilibrium, is still under thermochemical non-equilibrium. But larger time steps mean that the system is given enough time to reach equilibrium from its initial state. This final equilibrium state depends on the initial total energy that the system starts with, since we are in an adiabatic reactor case and this means that the total energy is conserved. For the chosen range of initial conditions, the initial total energy may be strongly dependent on the mass fraction of group-1, which is why we see this quantity to become influential at larger time steps.    
This is also apparent in the Sobol index plots of $Y_1$ where its initial value remains the most influential parameter throughout the chosen time domain.  
Additionally, $Y_2$ stands out as the most influenced by the model error parameter $\delta$, along with $T_2$, which is also moderately affected by 
$\delta$. 
This could be linked to the fact that as the energy partition parameter directly influences the grouping strategy, it has a dominant influence on the relaxation dynamics for the high-lying energy levels. 
These insights are more difficult to obtain with black-box machine learning surrogates.   

\subsection{Surrogate performance on test samples}
In this section, we assess the predictive capability of the surrogate model when used in time marching simulations for 0-D adiabatic reactor and 1-D shock case, where it is coupled with a fluid solver. 

Table \ref{tab:table_test_error} shows the percentage mean relative error calculated according to \cref{eq:per_rel_err} on 100 test samples taken from the range shown in table \ref{tab:table_range_ics} for the initial conditions and samples from the joint posterior of model error and parametric uncertainty.
\begin{equation}
    \text{\% rel. err} = 100 \times \frac{1}{N} \sum_{i=1}^N \frac{||g_i(\boldsymbol{\phi}, \delta, \boldsymbol{\kappa}_L, t) - \hat{g}_i(\boldsymbol{\phi}, \delta, \boldsymbol{\kappa}_L, t)||_2}{||g_i(\boldsymbol{\phi}, \delta, \boldsymbol{\kappa}_L, t)||_2}\label{eq:per_rel_err}
\end{equation}
In the above equation, $\hat{g}_i$ denotes the prediction from the surrogate at the $i$th input and $g_i$ denotes the true solution. 
The surrogate model demonstrates excellent predictive accuracy for all QoIs, with a maximum error below 10\%. The largest errors occur in $T_2$ and $Y_2$, quantities most affected by the model error parameter.
\begin{center}
\captionof{table}{Error on test samples}
\label{tab:table_test_error}
\begin{tabular}{|l|l|}
\hline
Quantity  &  \% rel. error\\
\hline
$T$ & 2.47\\
\hline
$T_1$ & 2.23\\
\hline
$T_2$ & 5.09\\
\hline
$Y_1$ & 2.43\\
\hline
$log_{10}(Y_2)$ & 8.39\\
\hline
\end{tabular}
\end{center}

\subsubsection{Zero dimensional adiabatic reactor}

\begin{figure}[H]
    \centering
    \subcaptionbox{\label{fig:adia_test_surr_T}}
    {\includegraphics[width=0.32\linewidth]{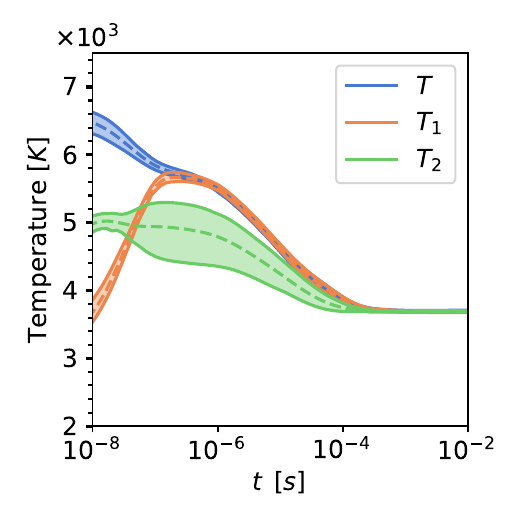}}
    \subcaptionbox{\label{fig:adia_test_surr_Y}}
    {\includegraphics[width=0.32\linewidth]{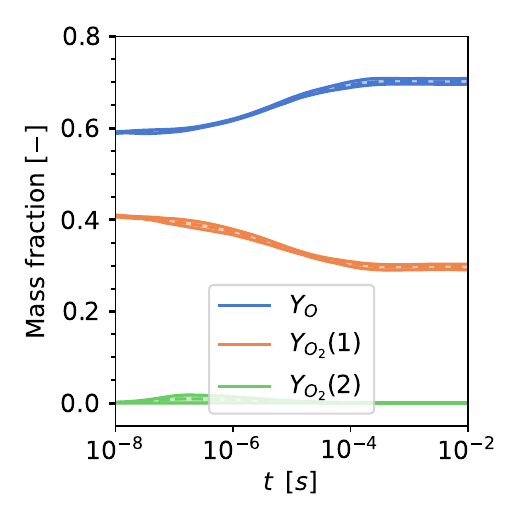}}
    \subcaptionbox{\label{fig:adia_test_surr_pop}}{\includegraphics[width=0.32\linewidth]{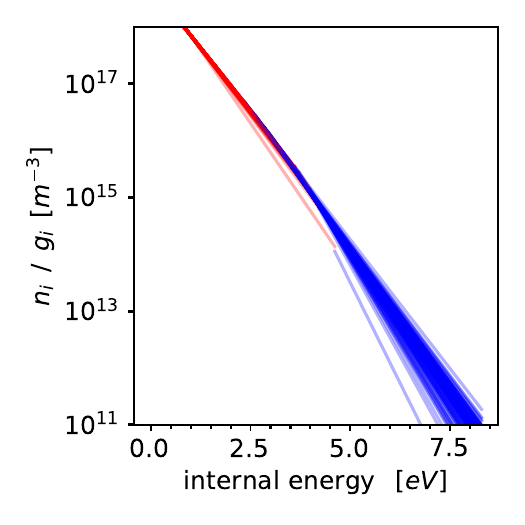}}
    \caption{Posterior predictive mean shown with dotted line along with 95\% prediction intervals for temperatures (a) and mass fractions (b) obtained with surrogate. Reconstructed population distribution (c) for 100 samples at $t = 3 \times 10^{-7} \,s$.} 
    \label{fig:adia_test_surr}
\end{figure}

\begin{figure}[H]
    \centering
    \subcaptionbox{\label{fig:adia_test_ode_T}}
    {\includegraphics[width=0.32\linewidth]{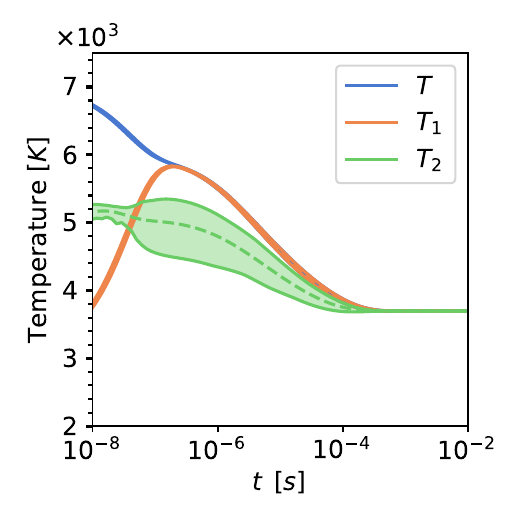}}
    \subcaptionbox{\label{fig:adia_test_ode_Y}}
    {\includegraphics[width=0.32\linewidth]{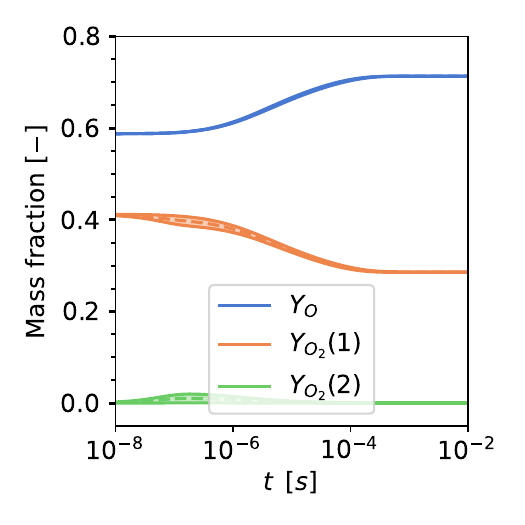}}
    \subcaptionbox{\label{fig:adia_test_ode_pop}}{\includegraphics[width=0.32\linewidth]{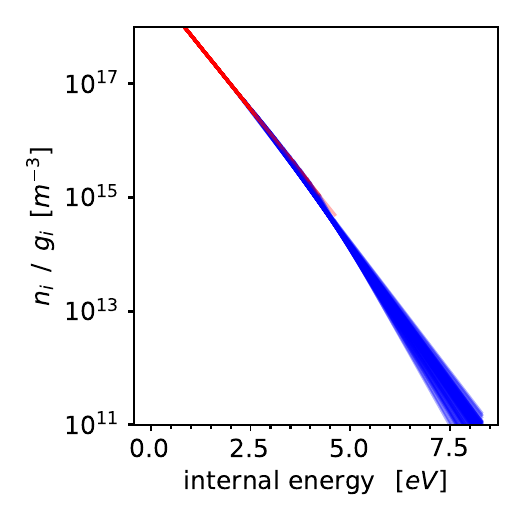}}
    \caption{Posterior predictive mean shown with dotted line along with 95\% prediction intervals for temperatures (a) and mass fractions (b) obtained with ODE solver. Reconstructed population distribution (c) for 100 samples at $t = 3 \times 10^{-7} \,s$.}
    \label{fig:adia_test_ode}
\end{figure}

\begin{figure}[H]
    \centering
    
    \subcaptionbox{\label{fig:adia_test_KDE_T}}
    {\includegraphics[width=0.33\linewidth]{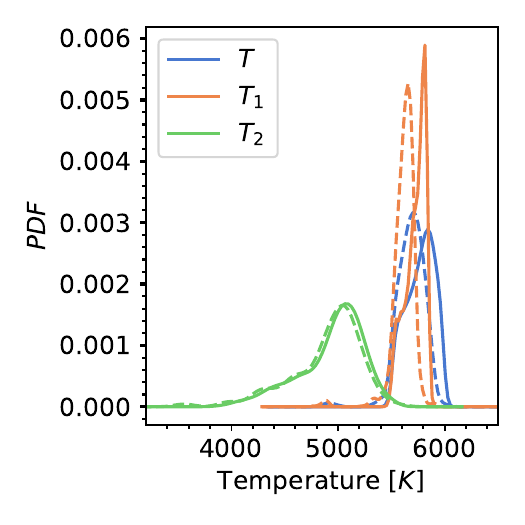}}
    \subcaptionbox{\label{fig:adia_test_KDE_Y}}
    {\includegraphics[width=0.33\linewidth]{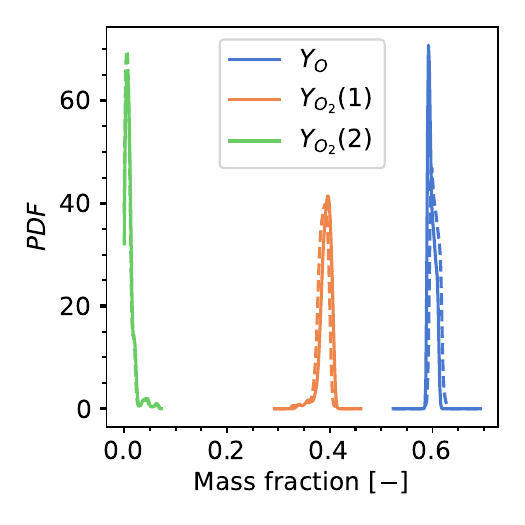}}
    \caption{PDFs of temperatures (a) and mass fractions (b) from t = $10^{-7}$ to $10^{-6} \,s$; solid lines indicate PDFs obtained from ODE solver, dashed lines indicate PDFs obtained with surrogate.}
    \label{fig:adia_test_KDE}
\end{figure}

Fig. \ref{fig:adia_test_surr_T} and Fig. \ref{fig:adia_test_surr_Y} show the posterior predictive distributions for temperatures and mass fractions, respectively, obtained from the surrogate model while Fig. \ref{fig:adia_test_ode_T} and Fig. \ref{fig:adia_test_ode_Y} show the predictive distributions obtained with an ODE solver for the test initial conditions:
$\rho_0 = 0.02 \, kg/m^3, T_0 = 6870 \, K, T_{10} = T_{20} = 3070 \, K, Y_{10} = 0.59, Y_{20} = 10^{-5}$. Predictions from both methods show excellent agreement. Moreover, the surrogate model remains stable over time, with no accumulation of error that might otherwise make time integration unstable. Achieving this level of accuracy required retaining enough KLE modes and targeting a low LOOCV error in PCE construction. However, as time progresses, the surrogate model accumulates some error, leading to slightly wider 95\% prediction intervals compared to the ODE solver. Increasing the surrogate model's accuracy—albeit at a higher computational cost—could mitigate this issue. 
The highest uncertainty is observed in $T_2$, while relatively smaller uncertainties propagate to the other QoIs, most likely due to the high influence of model error on the dynamics of group-2. 
The population reconstructions, which capture the impact of model error and parametric uncertainty on the STS distribution, obtained at $t=3 \times 10^{-7} \, s$, using \cref{eq:grp_recon}, for the surrogate and ODE solver are shown in Fig. \ref{fig:adia_test_surr_pop} and Fig. \ref{fig:adia_test_ode_pop} respectively. Both exhibit good agreement. We also see higher uncertainty in populations of high-lying energy levels, driven by the combined effects of model and parametric uncertainty, particularly due to strong dissociation from these states at the chosen time. 
Figs. \ref{fig:adia_test_KDE_T} and \ref{fig:adia_test_KDE_Y} present the summation of PDFs of temperatures and mass fractions from $t = 10^{-7}$ to $10^{-6} \, s$, respectively. The solid lines indicate the PDFs obtained with ODE solver, while the dotted lines indicate the PDFs obtained from the surrogate model.
We plot the PDFs across a range of the time domain rather than at a single point in time to avoid discrepancies that could arise from even slight shifts in the solution obtained with the two methods. This approach provides a clearer comparison.
Overall, we observe a good agreement between the two methods. 
While the surrogate model slightly underpredicts the temperatures $T$ and $T_1$, this can be improved by enhancing the surrogate's accuracy. Notably, despite the large uncertainty, the second group temperature $T_2$ is predicted very accurately by the surrogate.
 
\subsubsection{One dimensional shock}
In this section, we assess the predictive capability of our surrogate model when coupled with a flow solver  for predictions in 1-D normal shock as described in section \ref{sec:shock_goveq}. 
The test scenario has free stream conditions of $P_0 = 675 \, Pa, T_0 = 300 \, K, U_{0} = 5.895 \, km/s \, (M_{0} \approx 17)$. 

\begin{figure}[H]
    \centering
    
    \subcaptionbox{$t = 10^{-6}s$}{\includegraphics[width=0.32\linewidth]{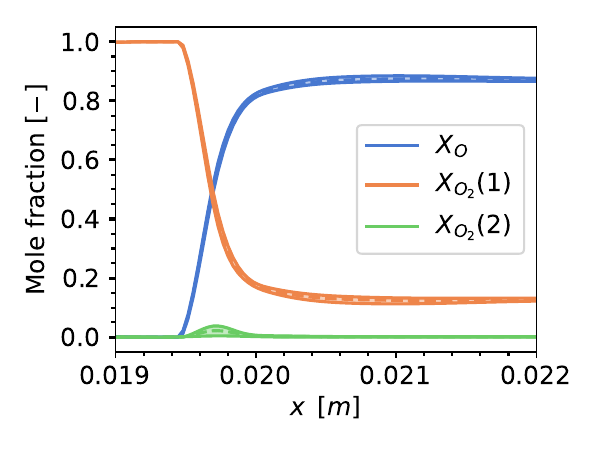}}
    \subcaptionbox{$t = 3 \times 10^{-6}s$}{\includegraphics[width=0.32\linewidth]{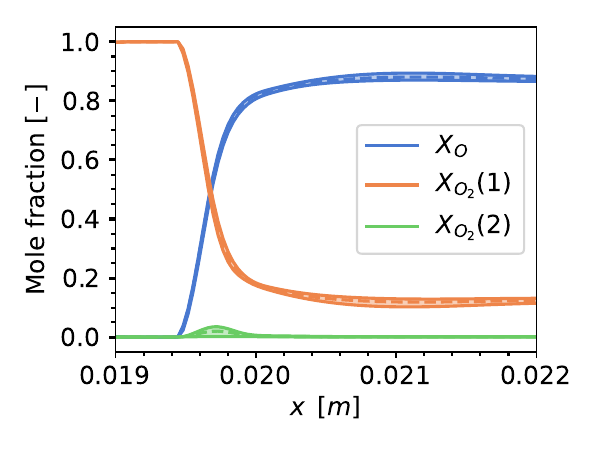}}
    \subcaptionbox{$t = 5 \times 10^{-6}s$}{\includegraphics[width=0.32\linewidth]{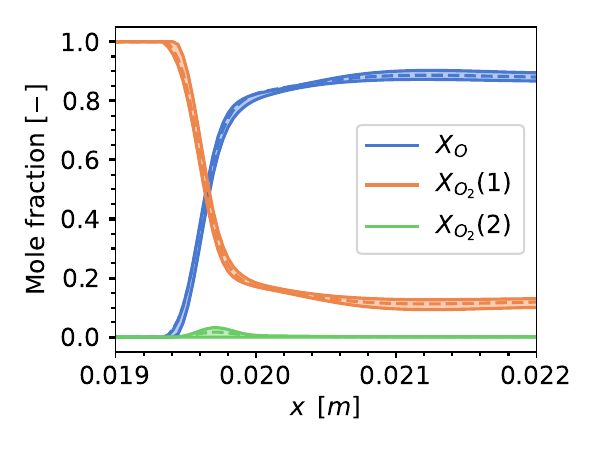}}
    \caption{Posterior predictive mean shown with dotted lines along with 95\% prediction intervals for mole fractions obtained with surrogate.}
    \label{fig:shock_test_surr_mole}
\end{figure}

\begin{figure}[H]
    \centering
    
    \subcaptionbox{$t = 10^{-6}s$}{\includegraphics[width=0.32\linewidth]{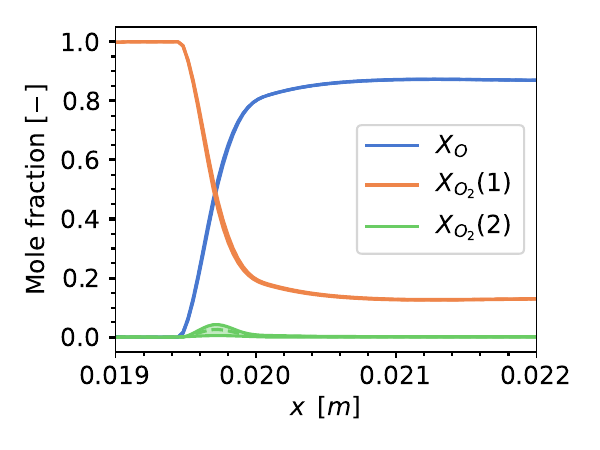}}
    \subcaptionbox{$t = 3 \times 10^{-6}s$}{\includegraphics[width=0.32\linewidth]{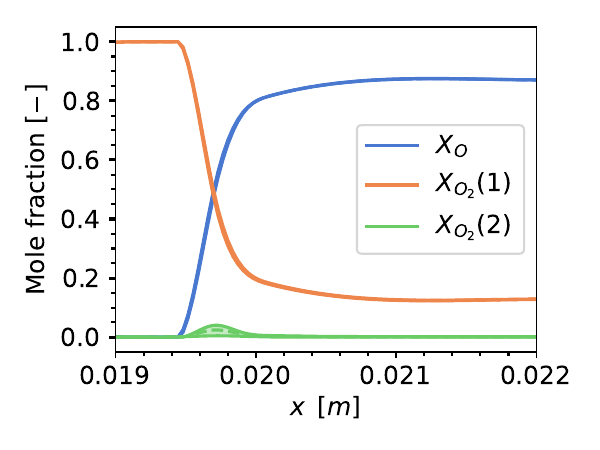}}
    \subcaptionbox{$t = 5 \times 10^{-6}s$}{\includegraphics[width=0.32\linewidth]{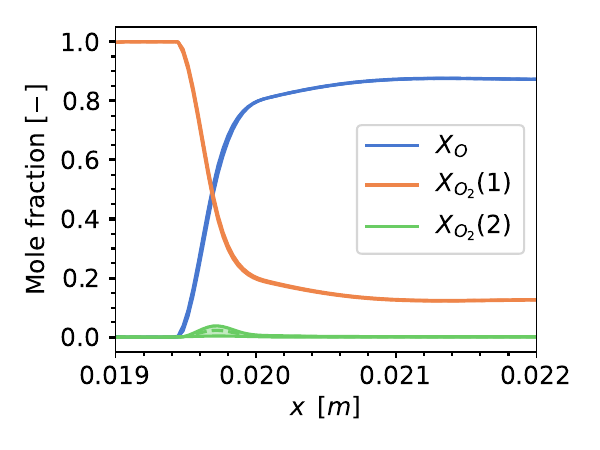}}
    \caption{Posterior predictive mean shown with dotted lines along with 95\% prediction intervals for mole fractions obtained with ODE solver.}
    \label{fig:shock_test_ode_mole}
\end{figure}

\begin{figure}[H]
    \centering
    
    \subcaptionbox{}{\includegraphics[width=0.32\linewidth]{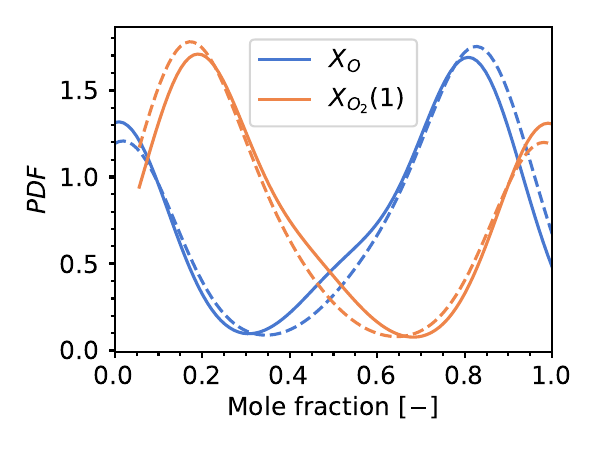}}
    \subcaptionbox{}{\includegraphics[width=0.32\linewidth]{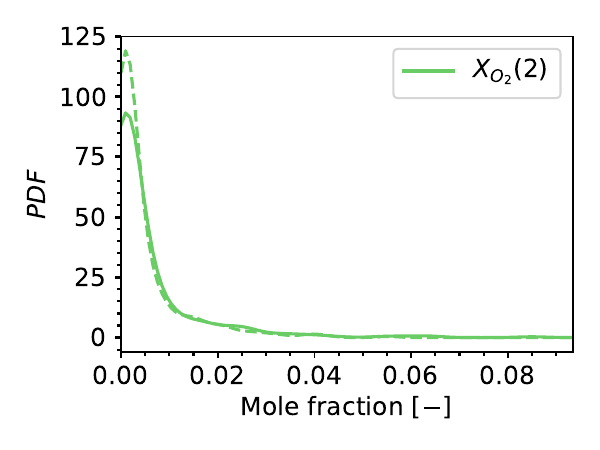}}
    \caption{PDFs of $X_O$, $X_{O_2}(1)$ (a) and $X_{O_2}(2)$ (b) between x = $0.019$ to $0.0205\,m$ at t = $5\times 10^{-6}\,s$; solid lines indicate PDFs obtained with ODE solver, dashed lines indicate PDFs obtained with surrogate.}
    \label{fig:KDEs_shock_mole_frac}
\end{figure}

Figures \ref{fig:shock_test_surr_mole} and \ref{fig:shock_test_ode_mole} present the species mole fraction predictions in shock reference frame, obtained using the surrogate model as a replacement for \cref{eq:react_1} and a conventional ODE solver for \cref{eq:react_1}, respectively, at three different time instances. Once again, we observe excellent agreement between the two approaches. As seen in Fig. \ref{fig:shock_test_surr_mole}, the surrogate model accumulates a small amount of error over time, which results in slightly wider prediction intervals. However, the surrogate model remains stable, as the error is not significant enough to cause numerical instabilities during time marching. Fig. \ref{fig:KDEs_shock_mole_frac} shows the PDFs of mole fractions between $x = 0.019$ to $0.0205 \, m$ at the last time instant, $t = 5 \times 10^{-6} \, s$. Solid lines represent the PDFs obtained using the ODE integrator, while dotted lines show those obtained with the surrogate. Despite a slight discrepancy due to the surrogate's error accumulation, the two solutions demonstrate very good agreement.    

\begin{figure}[H]
    \centering
    
    \subcaptionbox{$t = 10^{-6}s$}{\includegraphics[width=0.32\linewidth]{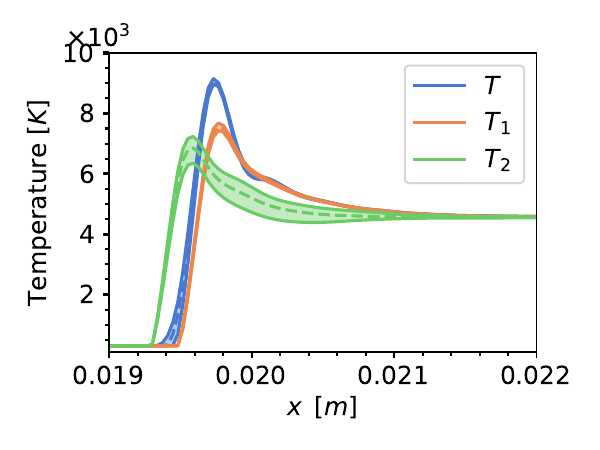}}
    \subcaptionbox{$t = 3 \times 10^{-6}s$}{\includegraphics[width=0.32\linewidth]{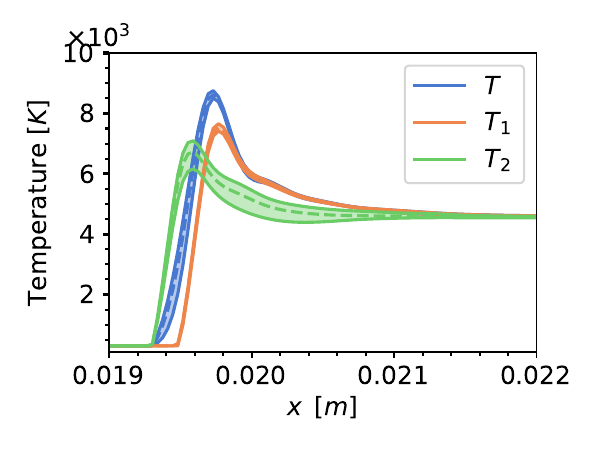}}
    \subcaptionbox{$t = 5 \times 10^{-6}s$}{\includegraphics[width=0.32\linewidth]{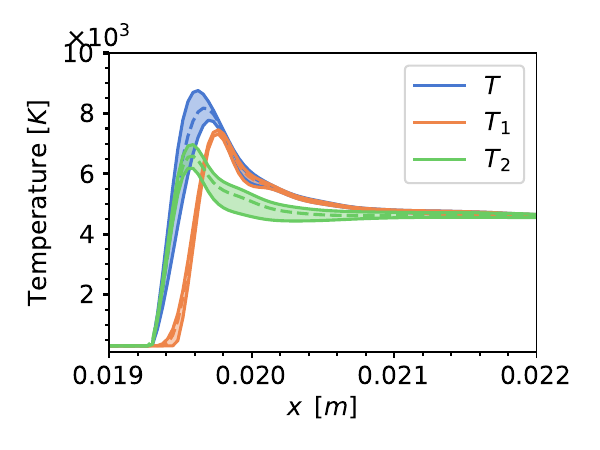}}
    \caption{Posterior predictive mean shown with dotted lines along with 95\% prediction intervals for temperatures obtained with surrogate.}
    \label{fig:shock_test_surr_T}
\end{figure}

\begin{figure}[H]
    \centering
    \subcaptionbox{$t = 10^{-6}s$}{\includegraphics[width=0.32\linewidth]{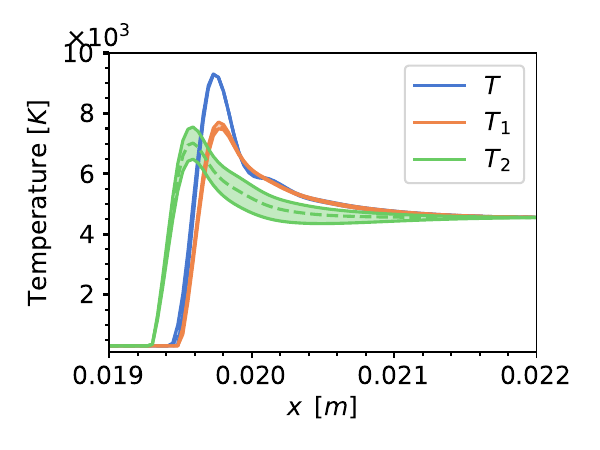}}
    \subcaptionbox{$t = 3 \times 10^{-6}s$}{\includegraphics[width=0.32\linewidth]{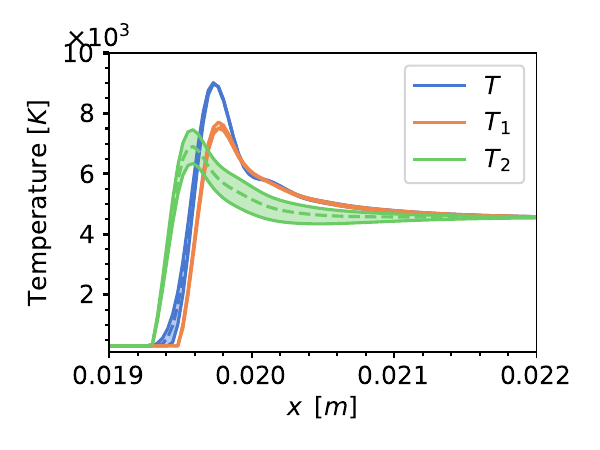}}
    \subcaptionbox{$t = 5 \times 10^{-6}s$}{\includegraphics[width=0.32\linewidth]{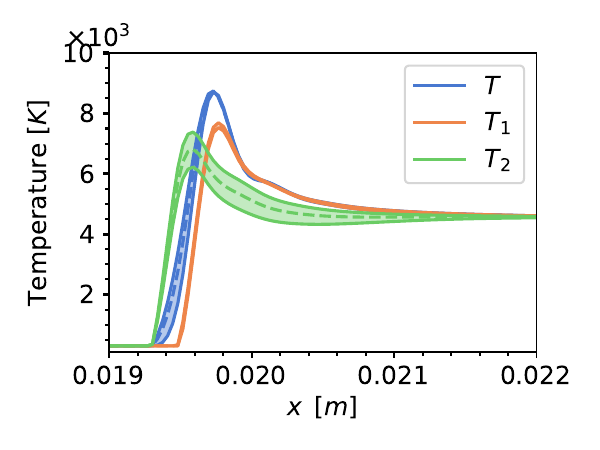}}
    \caption{Posterior predictive mean shown with dotted lines along with 95\% prediction intervals for temperatures obtained with ODE solver.}
    \label{fig:shock_test_ode_T}
\end{figure}

\begin{figure}[H]
    \centering
    
    \subcaptionbox{}{\includegraphics[width=0.32\linewidth]{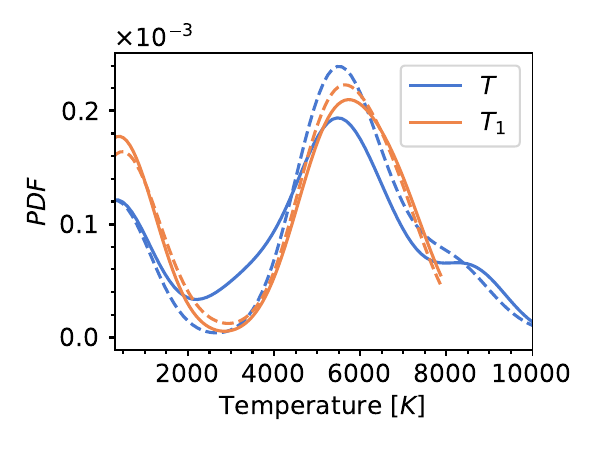}}
    \subcaptionbox{}{\includegraphics[width=0.32\linewidth]{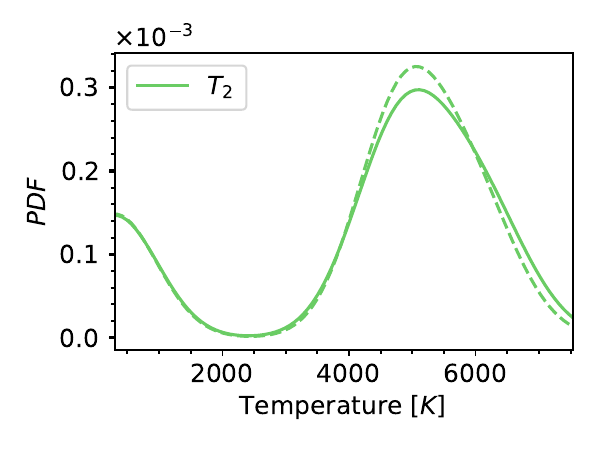}}
    \caption{PDFs of $T$, $T_1$ (a) and $T_2$ (b) between x = $0.019$ to $0.0205\,m$ at t = $5\times 10^{-6}\,s$; solid lines indicate PDFs obtained with ODE solver, dashed lines indicate PDFs obtained with surrogate.}
    \label{fig:KDEs_shock_T}
\end{figure}

Fig. \ref{fig:shock_test_surr_T} and Fig. \ref{fig:shock_test_ode_T} show the predictions in temperatures obtained from the surrogate and ODE solver, respectively. Again, we have a good agreement between the truth and surrogate predictions, with slightly wider prediction intervals with surrogate due to surrogate error propagation over time. We see the highest uncertainty in group-2 temperature $T_2$ as seen previously in the adiabatic reactor case. Fig. \ref{fig:KDEs_shock_T} compares the PDFs obtained with surrogate (dashed lines) and ODE solver (solid lines) between $x = 0.019$ to $0.0205 \, m$ at $t=5 \times 10^{-6} \, s$. In general, we see a good agreement for all three temperatures, with slightly more discrepancy in $T$ which is the most affected by surrogate model's error accumulation. 

These results demonstrate the effectiveness of combining KLE and PCE to accurately estimate the stochastic operator governing the evolution of chemical kinetics under model and parametric uncertainty. The surrogate did not provide a significant speed-up compared to the ODE solver, as the chemistry model considered is relatively small, with only three species. However, the primary objective was to construct a surrogate that delivers stable probabilistic predictions, which the results clearly illustrate. Surrogating larger systems will likely yield better speed improvements. Additionally, the proposed framework shows promise for multi-dimensional CFD simulations due to its straightforward structure and compatibility with fluid solvers. Multi-dimensional CFD will demand a wider range of initial conditions, which can be effectively addressed by dividing the domain into smaller parts and constructing surrogates for each segment.

\section{Conclusions}\label{sec:Conclusions}
In this work, we proposed a robust framework for quantifying model error and parametric uncertainty in coarse-graining models for non-equilibrium chemical kinetics. Model error was accounted for in a physically consistent manner by embedding stochasticity at its source within the chemistry model through a Polynomial Chaos Expansion. Bayesian inference was then used to learn the posterior distributions of model and parametric uncertainty based on data from a high-fidelity model.
To accelerate inference and enable probabilistic predictions, we proposed a surrogate modeling methodology that learns the operator governing the chemistry model by separating time dynamics from the influence of other input parameters. We utilized the Karhunen-Loève Expansion to obtain temporal modes, and we mapped the latent quantities of interest, represented by the KLE coefficients, to the corresponding inputs via PCE. This approach yielded a model that was not only stable for time integration and easy to couple with external fluid solvers but also retained the physical interpretability often lacking in many machine learning models.

We applied this methodology to the $\mathrm{O}_2$-$\mathrm{O}$ chemistry system, using a 2-bin energy-based model as the low-fidelity model and a 10-bin spectral clustering model as the high-fidelity reference to provide inference data. Uncertainty in grouping was accounted for by making the energy group partition parameter stochastic.
Our application of the surrogate model to 0-D adiabatic reactor and 1-D shock simulations demonstrated stable predictions over time, with quantified uncertainty due to model error and rate coefficient uncertainty, and achieved a maximum relative prediction error of below 10\%.

Future work will focus on accounting for other sources of error, such as uncertainty due to truncation in the reconstruction strategy. Additionally, the surrogate modeling could be enhanced by incorporating physical constraints, as suggested in the physics-informed PCE method \cite{novak2024physics}. Given that this study deals with a coupled system of ODEs, including physical constraints would involve formulating a joint optimization problem for PCE coefficients of all QoIs to ensure robust solutions. Furthermore, bootstrap resampling \cite{marelli2018active} could be employed to address issues related to finite sample sizes, given that regression was used to obtain the PCE coefficients.
Future work will also focus on surrogating larger chemistry systems with both model and parametric uncertainty, which would also lead to computational speed-up. This will require inference methods suited for large-scale inverse problems \cite{lieberman2010parameter,martin2012stochastic} and surrogate models that provide accurate derivative information while preserving physical constraints \cite{o2024derivative}.

\appendix

\section{Additional details}\label{sec:Appendix}

\subsection{Truncation of Karhunen Lo\`{e}ve Expansion}\label{sec:Appendix-KLE-truncation}

\begin{figure}[H]
    \centering
    \subcaptionbox{$T$}
    {\includegraphics[width=0.3\linewidth]{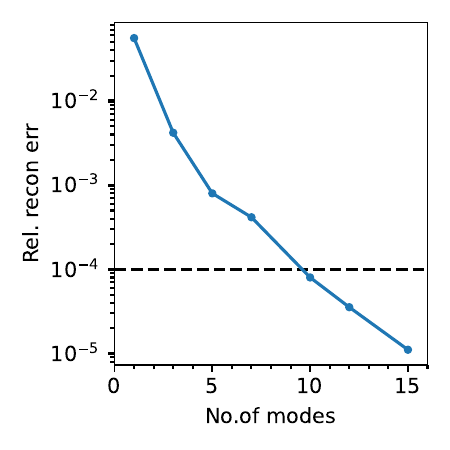}}
    \subcaptionbox{$T_1$}
    {\includegraphics[width=0.3\linewidth]{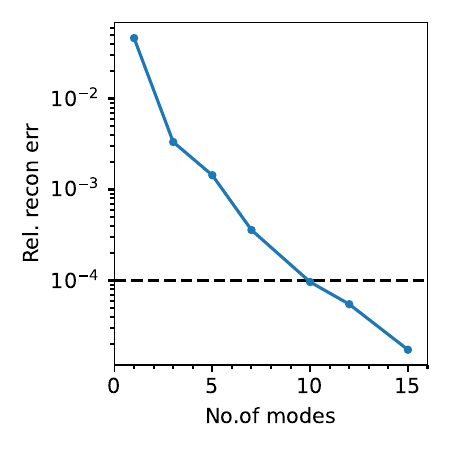}}
    \subcaptionbox{$T_2$}
    {\includegraphics[width=0.3\linewidth]{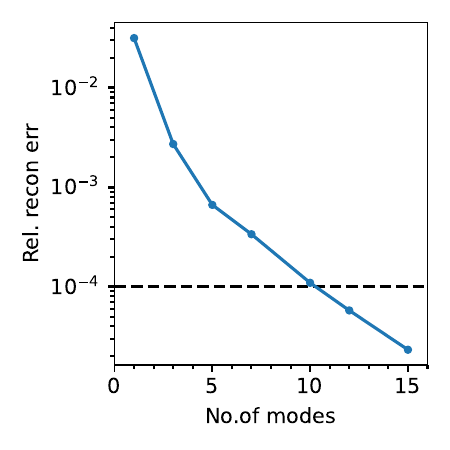}}
    \subcaptionbox{$Y_1$}
    {\includegraphics[width=0.3\linewidth]{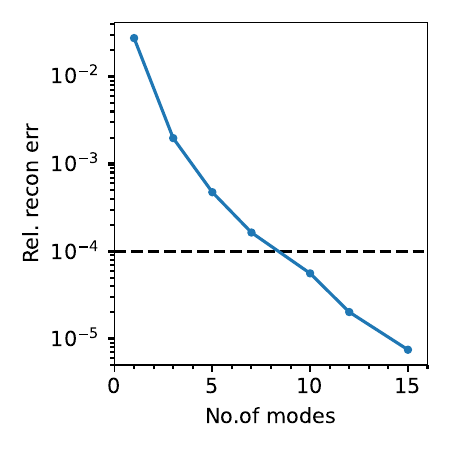}}
    \subcaptionbox{$Y_2$}
    {\includegraphics[width=0.3\linewidth]{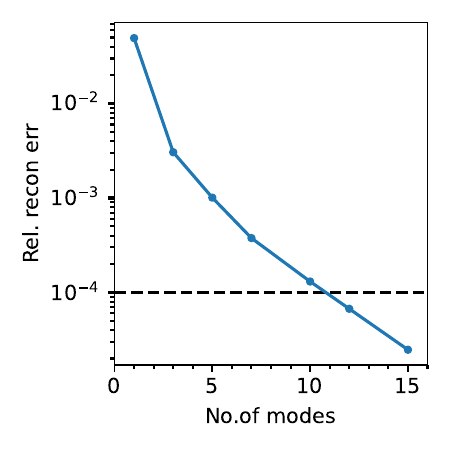}}
    \caption{Mean relative reconstruction error as a function of number of KLE modes. The number of modes is chosen so that the reconstruction error is below $10^{-4}$, shown by the dotted black line}
    \label{fig:rel_error_KLE}
\end{figure}

The number of modes for each QoI is chosen so that the mean relative reconstruction error across the training samples obtained is below $10^{-4}$:
\begin{equation}
    \text{rel. err}(P) = \frac{1}{N_{train}} \sum_{i=1}^{N_{train}} \frac{||g_i(\boldsymbol{\phi}, \delta, \boldsymbol{\kappa}_L, t) - \hat{g}^{P}_i(\boldsymbol{\phi}, \delta, \boldsymbol{\kappa}_L, t)||_2}{||g_i(\boldsymbol{\phi}, \delta, \boldsymbol{\kappa}_L, t)||_2}\label{eq:mean_rel_err}
\end{equation}
where $\hat{g}^{P}_i$ is used to represent the reconstruction of QoI $g$ at the $i$th training input with $P$ modes. Figure \ref{fig:rel_error_KLE} shows this error as a function of the number of modes for each QoI. 

\subsection{Accuracy of Polynomial Chaos Expansion}

\begin{figure}[H]
    \centering
    \subcaptionbox{\label{fig:joint_dist_KLE_coef_true}}
    {\includegraphics[width=0.49\linewidth]{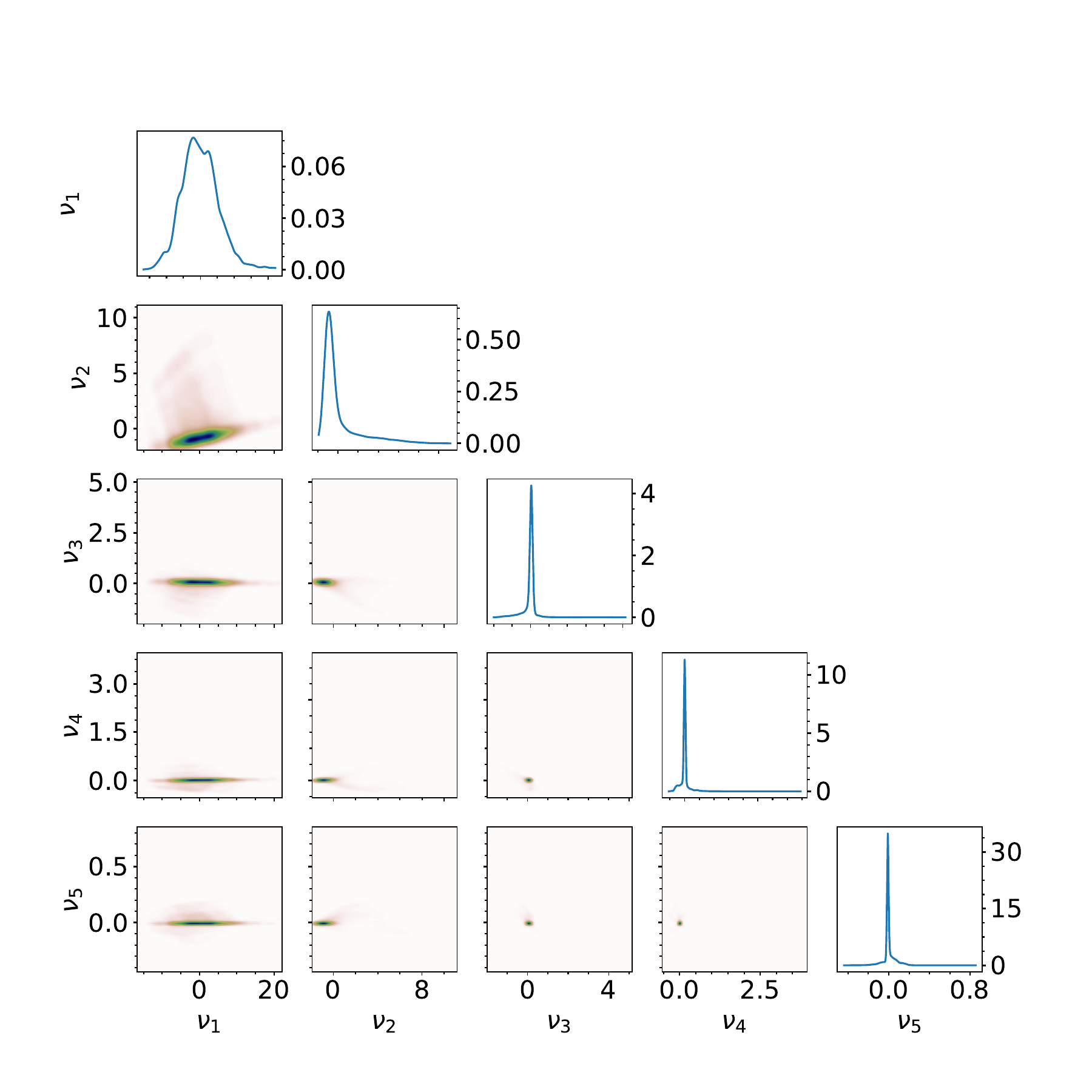}}
    \subcaptionbox{\label{fig:joint_dist_KLE_coef_surr}}
    {\includegraphics[width=0.49\linewidth]{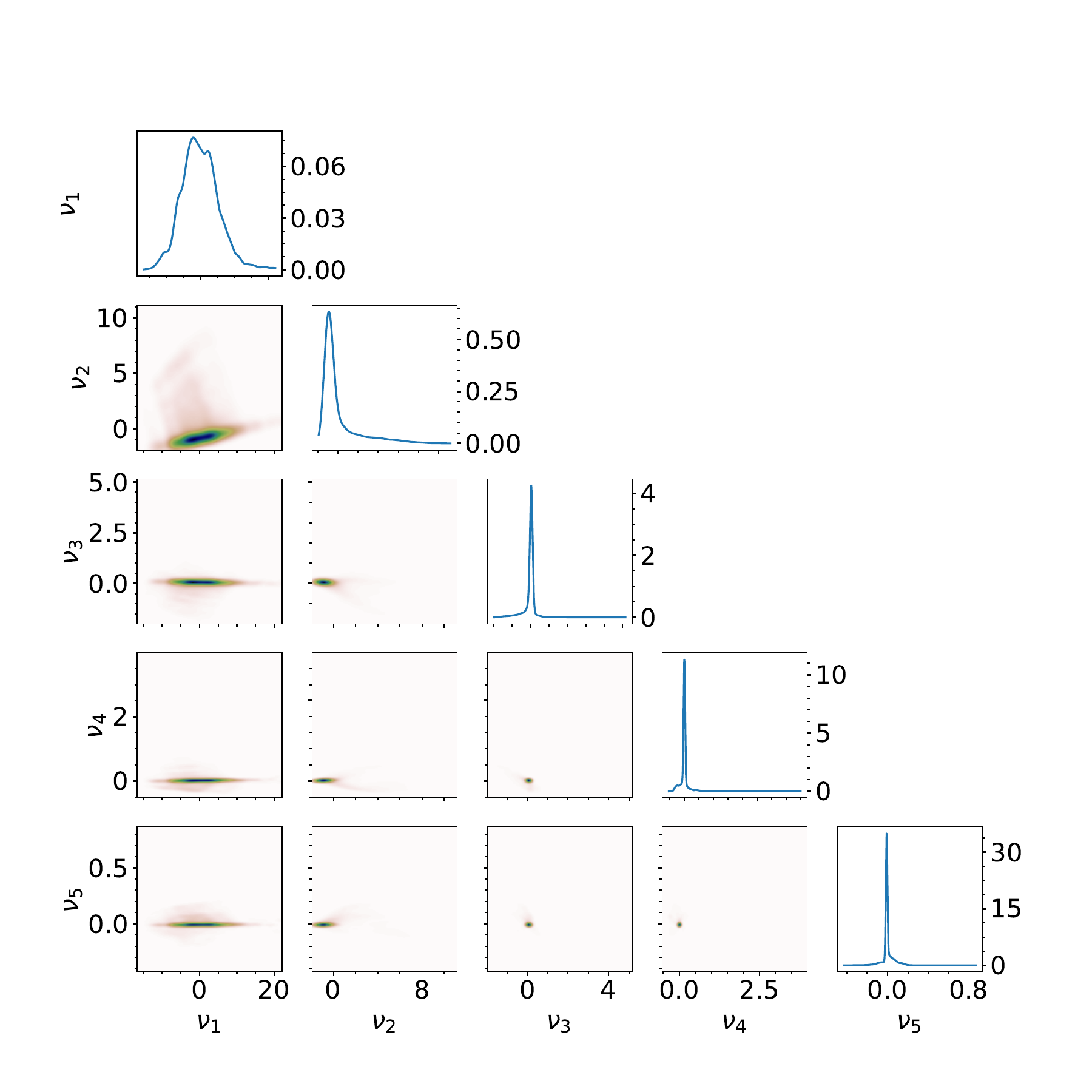}}
    \caption{Joint distribution of top five dominant KLE coefficients obtained with training data (a) and PCEs prediction on training inputs (b) for QoI $Y_2$}
    \label{fig:joint_dist_KLE_coef}
\end{figure}

LAR was used to obtain the PCE coefficients, where the PCE order was adaptively increased to 10, until a variance normalized LOOCV error of $10^{-6}$ was achieved. Table \ref{tab:table_LOOCV}
 shows the LOOCV values along with PCE order achieved by the most dominant mode of each QoI. It can be seen that except for $Y_1$, no other QoI was able to achieve an error below the specified target of $10^{-6}$ even at the highest order. This suggests targeting orders higher than 10 for these QoIs to achieve LOOCV below $10^{-6}$.  
\begin{center}
\captionof{table}{Leave-one-out variance normalized cross validation error for the most dominant mode}
\label{tab:table_LOOCV}
\begin{tabular}{|l|l|l|}
\hline
Quantity  & PCE order & LOOCV\\
\hline
$T$ & 10 & $0.227 \times 10^{-5}$\\
\hline
$T_1$ & 10 & $0.259 \times 10^{-5}$\\
\hline
$T_2$ & 10 & $0.727 \times 10^{-5}$\\
\hline
$Y_1$ & 5 & $0.731 \times 10^{-6}$\\
\hline
$Y_2$ & 10 & $0.140 \times 10^{-4}$\\
\hline
\end{tabular}
\end{center}
Figs \ref{fig:joint_dist_KLE_coef_true} and \ref{fig:joint_dist_KLE_coef_surr} show the joint distribution of five dominant KLE coefficients obtained with the training samples and that obtained with the PCE predictions on the training inputs for QoI $Y_2$, respectively. We note that each coefficient is multiplied by the square root of its corresponding eigenvalue, and hence the variance in each of the marginal distributions is not 1. Despite having the highest LOOCV among all QoIs, we see that the joint distribution is accurately captured by the constructed surrogate model. Moreover, the distributions are highly non-Gaussian, which led to high orders in the PCEs.   

\subsection{Model error embedding incorporated as random Polynomial Chaos Expansion coefficients}

\begin{figure}[H]
    \centering
    {\includegraphics[width=0.59\linewidth]{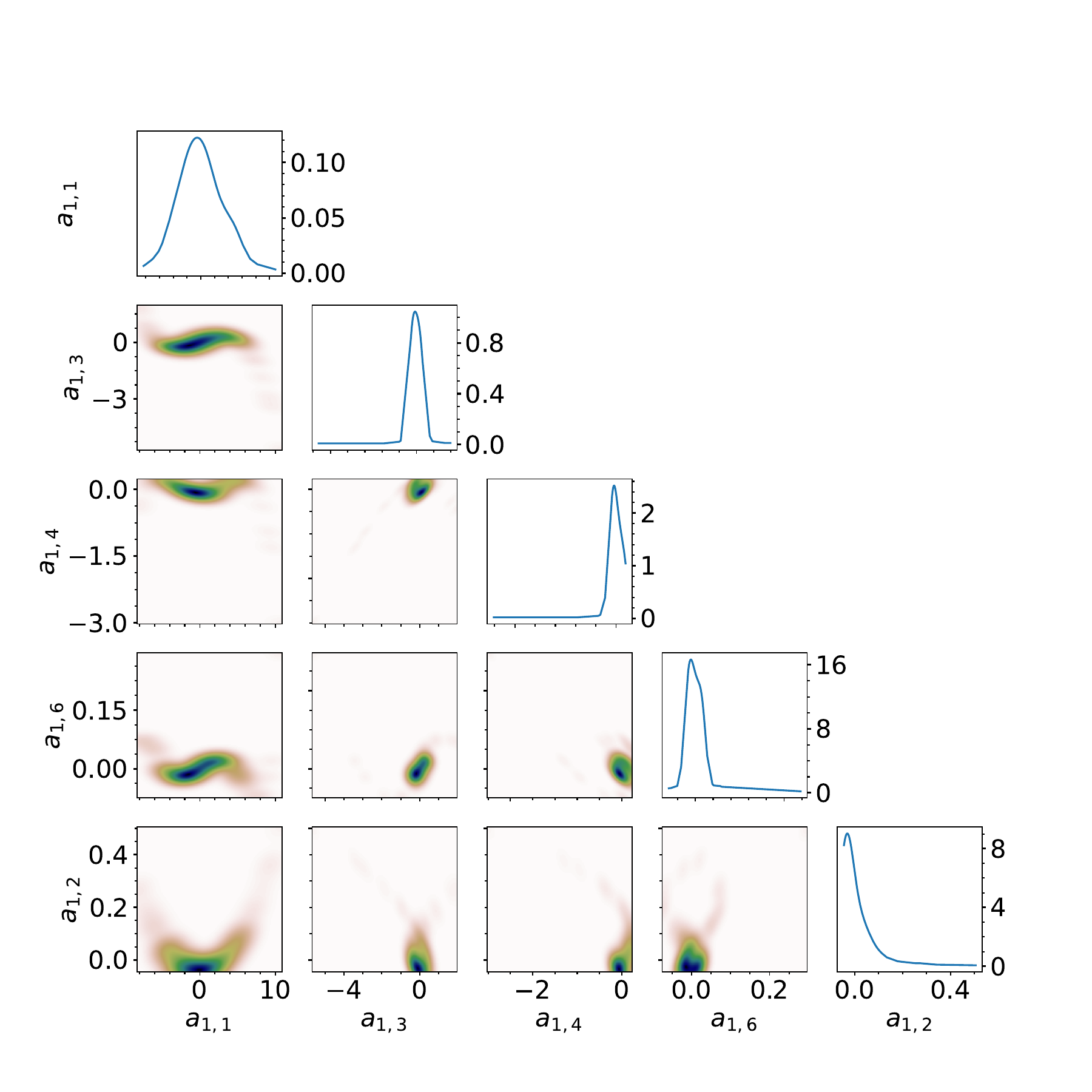}}
    \caption{Joint distribution of top five dominant PCE coefficients for the 1st mode of $Y_2$. $j$ subscript in $a_{1,j}$ denotes the degree of $\boldsymbol{\xi}_{\boldsymbol{\delta}}$ in $\psi_{\boldsymbol{\alpha}_{\boldsymbol{\delta}}}(\boldsymbol{\xi}_{\boldsymbol{\delta}})$}
    \label{fig:pce_rand_coef_Y2}
\end{figure}

As detailed in section \ref{sec:rand_pce_coef}, the PCE basis in model error parameter, $\psi_{\boldsymbol{\alpha}_{\boldsymbol{\delta}}}(\boldsymbol{\xi}_{\boldsymbol{\delta}})$, can be incorporated into the PCE coefficients, thus obtaining distributions of PCE coefficients that incorporate the model error. The joint distribution of top five dominant PCE coefficients is shown in Fig. \ref{fig:pce_rand_coef_Y2}, for the most dominant mode of $Y_2$. It can be seen that the model error is highly non-Gaussian, and the structure of the surrogate model enables us to capture this non-Gaussian distribution accurately. 

\section*{Acknowledgements}

This research was supported by ONR under the MURI grant no. N00014-21-1-2475. The views and conclusions contained herein are those of the authors and should not be interpreted as necessarily representing the official policies or endorsements, either expressed or implied, of the ONR or the US government. 

\bibliographystyle{elsarticle-num} 


\end{document}